\documentclass[manuscript]{acmart}

\AtBeginDocument{%
  \providecommand\BibTeX{{%
    \normalfont B\kern-0.5em{\scshape i\kern-0.25em b}\kern-0.8em\TeX}}}

\usepackage{algorithmicx}
\usepackage[noend]{algpseudocode}
\usepackage{amsmath,amsthm} 
\usepackage{array}
\usepackage{caption}
\usepackage{subcaption}
\usepackage{graphicx}
\usepackage{xcolor,colortbl}

\newcommand{\rt}[1]{\textcolor{red}{#1}}

\newcommand{\ct}[1]{{\color{blue}{#1}}}

\newcommand{\ot}[1]{\textcolor{orange}{#1}}
\newcommand{\sep}[1]{\textcolor{brown}{#1}}
\newcommand{\vl}[1]{\textcolor{violet}{#1}}

\usepackage{epstopdf}
\usepackage{setspace}
\usepackage{romannum}
\usepackage{enumerate}
\usepackage{ragged2e}
\usepackage{multirow}
\usepackage{tabularx,tabulary}
\usepackage{nicefrac}		
\usepackage{todonotes}
\usepackage[flushleft]{threeparttable}
\usepackage[ruled,vlined]{algorithm2e}
\usepackage{hyperref}
\SetKwRepeat{Do}{do}{while}
\usepackage{rotating}
\usepackage{tabularx,ragged2e,booktabs}
\usepackage[justification=centering]{caption}
\usepackage{listings}
\usepackage{pifont}
\usepackage[most]{tcolorbox}
\definecolor{backcolour}{rgb}{0.95,0.95,0.91}
\usepackage{lstlinebgrd}
\lstdefinestyle{customc}{
escapechar=?,
  abovecaptionskip=-5pt,
  breaklines=true,
  numbers=left, 
  floatplacement=t,
  frame = TRBL,
  xleftmargin=\parindent,
  language=C,
  showstringspaces=false,
  basicstyle=\footnotesize\ttfamily,
  keywordstyle=\bfseries\color{green!40!black},
  commentstyle=\itshape\color{purple!40!black},
  identifierstyle=\color{blue},
  stringstyle=\color{orange},
}
\newcolumntype{C}[1]{>{\Centering}m{#1}}

\usepackage{mathtools}
\usepackage{verbatim}

\usepackage{soul}
\usepackage{xcolor}

\usepackage{acronym}
\acrodef{GreyConE}{fuzzing and concolic execution}
\acrodef{HLS}{high-level synthesis}
\acrodef{IP}{intellectual property}
\acrodef{DUT}{design-under-test}
\acrodef{COTS}{commercial-off-the-shelf}
\acrodef{SoC}{system-on-chip}
\acrodef{CGF}{coverage-guided greybox fuzzing}
\acrodef{AFL}{American fuzzy lop}
\acrodef{S2E}{symbolic executer}
\acrodef{HDL}{hardware description language}
\acrodef{SAT}{Satisfiability}
\acrodef{AFL-SHT}{AFL-SHT}
\acrodef{SCT-HTD}{SCT-HTD}
\acrodef{BCOV}{Branch coverage}
\acrodef{LCOV}{Line coverage}
\acrodef{FCOV}{function coverage}
\acrodef{RTL}{register-transfer level}
\acrodef{IR}{intermediate representation}
\acrodef{SHT}{Synthesizable Hardware Trojan}

\usepackage{pifont}
\newcommand{\cmark}{\ding{51}}%
\newcommand{\xmark}{\ding{55}}%

\hypersetup{
    colorlinks,
    linkcolor={red!50!black},
    citecolor={blue!50!black},
    urlcolor={red!50!black}
}
\usepackage{comment}

\begin{document}



\title{Scalable Test Generation to Trigger Rare Targets in High-Level Synthesizable IPs for Cloud FPGAs}


\author{Mukta Debnath}
\email{mukta_t@isical.ac.in}
\affiliation{%
  \institution{Indian Statistical Institute}
  \city{Kolkata}
  \country{India}}
  
\author{Animesh Basak Chowdhury}
\email{abc586@nyu.edu}
\affiliation{%
  \institution{New York University}
  \state{New York}
  \country{USA}
}

\author{Debasri Saha}
\email{debasri_cu@yahoo.in}
\affiliation{%
  \institution{A.K. Chowdhury School of IT,University of Calcutta}
  \city{Kolkata}
  \country{India}
}

\author{Susmita Sur-Kolay}
\email{ssk@isical.ac.in}
\affiliation{%
  \institution{Indian Statistical Institute}
  \city{Kolkata}
  \country{India}}



\begin{abstract}
High-Level Synthesis (HLS) has transformed the development of complex Hardware IPs (HWIP) by offering abstraction and configurability through languages like SystemC/C++, particularly for Field Programmable Gate Array (FPGA) accelerators in high-performance and cloud computing contexts.
These IPs can be synthesized for different FPGA boards in cloud, offering compact area requirements and enhanced flexibility.
HLS enables designs to execute directly on ARM processors within modern FPGAs without the need for Register Transfer Level (RTL) synthesis, thereby conserving FPGA resources. 
While HLS offers flexibility and efficiency, it also introduces potential vulnerabilities such as the presence of hidden circuitry, including the possibility of hosting hardware trojans within designs. In cloud environments, these vulnerabilities pose significant security concerns such as leakage of sensitive data, IP functionality disruption and hardware damage, necessitating the development of robust testing frameworks.
This research presents an advanced testing approach for HLS-developed cloud IPs, specifically targeting hidden malicious functionalities that may exist in rare conditions within the design. The proposed method leverages selective instrumentation, combining greybox fuzzing and concolic execution techniques to enhance test generation capabilities. Evaluation conducted on various HLS benchmarks, possessing characteristics of FPGA-based cloud IPs with embedded cloud related threats, demonstrates the effectiveness of our framework in detecting trojans and rare scenarios, showcasing improvements in coverage, time efficiency, memory usage, and testing costs compared to existing methods.
\end{abstract}



\maketitle

\section{Introduction}
\label{label:intro}
In recent years, hardware design practices have shifted towards the widespread adoption of high-level synthesis (HLS) frameworks like SystemC and synthesizable C/C++~\cite{rosetta}. These frameworks enable the creation of complex hardware IPs (HWIPs) in high-level languages, facilitating implementation on Field Programmable Gate Arrays (FPGAs)~\cite{novel_simulation}. HLS streamlines hardware development through automated design space exploration and functional verification, offering higher abstraction compared to low-level RTL languages. When deploying FPGA-based solutions in public clouds, a myriad of considerations related to HWIPs come into play. IPs may be sourced from vendors or developed in-house by cloud service providers (CSPs). However, reliance on third-party IP cores (3PIPs) introduces security risks if the source is not fully trusted~\cite{security_of_cloud}. 

Moreover, the adoption of HLS introduces a vulnerability where attackers can clandestinely insert bug functionalities, such as hardware trojans, at a higher abstraction level within the design. On cloud-based FPGAs, trojans pose significant risks, potentially leaking sensitive information, inducing malfunctions in hardware design, tampering with computations or data, or maliciously altering underlying hardware~\cite{lifeofbitstream, security_of_cloud, trust-in-fpga}. Techniques like those described in~\cite{trojan-insertion} have been developed to detect trojans in FPGA designs at the IP and synthesis levels. These techniques rely on the principle that trojans are triggered rarely and can be identified through design profiling and simulation to probe for infrequently activated logic. Therefore, verifying the security aspects of HLS IPs is crucial for CSPs to ensure accurate functionality. Security testing of HLS IP cores tailored for cloud-based FPGAs is essential to identify and mitigate vulnerabilities, thereby maintaining a robust and secure foundation for IP cores deployed in public clouds.

Within the scope of this study, we introduce a scalable test-generation framework with the objective of triggering a potentially malicious functionality embedded in designs associated with third-party HLS IPs intended for FPGAs, as acquired by the CSP. Our framework, termed $GreyConE+$ leverages selective instrumentation techniques and builds upon two core testing techniques—fuzzing~\cite{afl} and concolic execution~\cite{sen2007concolic}—where each method is strategically employed to mitigate the limitations inherent in the other. This framework represents a continuation and enhancement of our earlier research on test generation for high-level languages, known as \textit{GreyConE}~\cite{greycone}. Through the integration of these advanced techniques, \textit{GreyConE+} offers a scalable and effective solution for identifying potential vulnerabilities within the rare conditions found in HLS IPs tailored for FPGAs hosted by CSPs. Our contributions in this study encompass the following key aspects:

\begin{enumerate}
\item We present the \textit{GreyConE+} framework designed to enhance the security of third-party HWIPs deployed on cloud-based FPGAs.
\item We introduce selective instrumentation method tailored for complex design IPs. This approach effectively targets security vulnerabilities that may be concealed within rare scenarios in designs, identified through random testing.
\item  \textit{GreyConE+} integrates greybox fuzzing and concolic execution techniques as guided test generation frameworks. These operate on partially instrumented designs with rare targets, offering a complementary solution to standalone approaches and mitigating associated challenges.
\item Leveraging \textit{GreyConE+} and our previous framework, \textit{GreyConE}~\cite{greycone}, we utilize reference model-based testing~\cite{model-based-testing} to uncover trojans concealed within compromised IPs on cloud-based FPGAs. This empowers CSPs to perform testing for the IPs on their own platform.
\item Our evaluation encompasses a diverse range of SystemC/C++ benchmark HLS IPs~\cite{scbench, s2cbench, s3cbench_benchmark}, including the ML-based Rosetta suite~\cite{rosetta} optimized for FPGAs. We assess \textit{GreyConE+}'s effectiveness in addressing rare scenarios and detecting trojan-triggered performance anomalies.
Key factors considered in our evaluation included improvements in time efficiency, target coverage, and reductions in testing costs related to memory usage and the number of generated test cases.
\end{enumerate}

The rest of the paper is organized as follows: Section~\ref{sec:background} outlines the background and the prior related works. Section~\ref{sec:motivation} discusses the motivation behind the work. Our assumed threat model is presented in Section~\ref{sec:threat-model}. In Section~\ref{sec:framework}, we propose our \textit{GreyConE+} framework and show the efficacy of results in Section~\ref{sec:results}. Section~\ref{analysis} presents an empirical analysis of the results and concluding remarks appear in Section~\ref{sec:conclusion}.

\begin{table}[!h]
\centering
\caption{Works on Security Testing of Hardware IPs}
\resizebox{\textwidth}{!}{
\begin{tabular}{@{}ccccc@{}}
\toprule \toprule
\textbf{Work} & \textbf{Abstraction level} & \textbf{Technique used} & \textbf{Benchmarks} & \textbf{Targets rare Triggers} \\
\midrule \midrule
\multirow{2}{*}{Saha \textit{et al.}~\cite{saha2015improved}} & \multirow{2}{*}{Gate level} & \multirow{2}{*}{Genetic algorithm + SAT formulation} & \multirow{2}{*}{ISCAS85, ISCAS89} & \multirow{2}{*}{\xmark} \\
\multirow{2}{*}{Chowdhury \textit{et al.}~\cite{bchowdhury2018}} & \multirow{2}{*}{Gate level} & \multirow{2}{*}{ATPG binning + SAT formulation} & \multirow{2}{*}{ISCAS85, ISCAS89, ITC99} & \multirow{2}{*}{\xmark} \\
\multirow{2}{*}{Huang \textit{et al.}~\cite{mers}} & \multirow{2}{*}{Gate level} & \multirow{2}{*}{Guided ATPG} & \multirow{2}{*}{ISCAS85, ISCAS89} & \multirow{2}{*}{\xmark} \\
\multirow{2}{*}{Liu \textit{et al.}~\cite{lyu2021maxsense}} & \multirow{2}{*}{Gate level} & \multirow{2}{*}{Genetic algorithm + SMT formulation} & \multirow{2}{*}{TrustHub~\cite{trusthub}} & \multirow{2}{*}{\xmark} \\
\multirow{2}{*}{Ahmed \textit{et al.}~\cite{iccd2017_concolic}} & \multirow{2}{*}{Register transfer level} & \multirow{2}{*}{Concolic testing} & \multirow{2}{*}{TrustHub~\cite{trusthub}} & \multirow{2}{*}{\cmark} \\
\multirow{2}{*}{Ahmed \textit{et al.}~\cite{itc2018_concolic}} & \multirow{2}{*}{Register Transfer Level} & \multirow{2}{*}{Greedy concolic testing} & \multirow{2}{*}{TrustHub~\cite{trusthub}} & \multirow{2}{*}{\cmark} \\
\multirow{2}{*}{Lyu \textit{et al.}~\cite{date2019_concolic}} & \multirow{2}{*}{Register Transfer Level} & \multirow{2}{*}{Parallelism + concolic testing} & \multirow{2}{*}{TrustHub~\cite{trusthub}}  & \multirow{2}{*}{\cmark} \\
\multirow{2}{*}{Veerana \textit{et al.}~\cite{s3cbench}} & \multirow{2}{*}{HLS/SystemC} & \multirow{2}{*}{Property checking} & \multirow{2}{*}{S3C~\cite{s3cbench_benchmark}} & \multirow{2}{*}{\cmark} \\
\multirow{2}{*}{Le \textit{et al.}~\cite{fuzzSystemC}} & \multirow{2}{*}{HLS/SystemC} & \multirow{2}{*}{Guided greybox fuzzing} & \multirow{2}{*}{S3C~\cite{s3cbench_benchmark}} & \multirow{2}{*}{\xmark} \\
\multirow{2}{*}{Bin \textit{et al.}~\cite{symbolicSystemC}} & \multirow{2}{*}{HLS/SystemC} & \multirow{2}{*}{Selective concolic testing} & \multirow{2}{*}{S3C~\cite{s3cbench_benchmark}} & \multirow{2}{*}{\xmark} \\
\multirow{2}{*}{Vafaei \textit{et al.}~\cite{symba}} & \multirow{2}{*}{C-level(from RTL)} & \multirow{2}{*}{Symbolic Execution} & \multirow{2}{*}{TrustHub~\cite{trusthub}} & \multirow{2}{*}{\cmark}\\
\multirow{2}{*}{\textit{GreyConE~\cite{greycone}}} & \multirow{2}{*}{HLS/SystemC,C/C++} & \multirow{2}{*}{Greybox Fuzzing(GF) + Concolic Execution(CE)} & \multirow{2}{*}{S3C~\cite{s3cbench_benchmark},S2C~\cite{s2cbench}} & \multirow{2}{*}{\xmark} \\
\multirow{2}{*}{\textbf{\textit{GreyConE+(Ours)}}} & \multirow{2}{*}{HLS/SystemC,C/C++} & \multirow{2}{*}{Selective Instrumentation with GF + CE} & \multirow{2}{*}{S3C~\cite{s3cbench_benchmark}, S2C~\cite{s2cbench},SC~\cite{scbench},Rosetta~\cite{rosetta}} & \multirow{2}{*}{\cmark} \\
&&&& \\ \bottomrule
\end{tabular}
}
\label{table:priorWork}
\end{table}

\section{Background}
\label{sec:background}

\subsection{Overview of Cloud IPs in HLS with malicious circuits}

In cloud environments with FPGA deployments for hardware acceleration, IPs are meticulously designed or optimized. These Cloud IPs, implemented using netlist, Hardware Description Language (HDL) or HLS, often incorporate specialized hardware accelerators to enhance performance in tasks like machine learning, data analytics, and cryptography. HLS techniques are increasingly common in cloud FPGA setups, enabling hardware designers to express functionality in higher-level languages such as SystemC, C or C++, which are then synthesized into FPGA configurations. Due to the critical role of security in cloud environments, it is imperative that these IPs are devoid of any malicious circuitry, aiming to enhance the overall security stance of FPGA-based solutions within the cloud. The presence of malicious functionalities like Trojans in IP designs introduces potential threats, including the risk of IP malfunctions, leakage of sensitive information, and potential damage to underlying hardware~\cite{s3cbench, farimah2021}. Vulnerabilities within third party HLS IPs might arise during the design phase, presenting similarities to security challenges encountered in untrusted electronics supply chains and the insertion of Trojans in pre-silicon hardware. Security researchers have proposed numerous ways to test and validate cloud-based IPs~\cite{HT3PIP,FPGAHT,AttacksFPGA,trust-in-fpga}.

\subsection{Security Testing of HWIPs}
\label{label:background_secTesting}
In the software community, security testing is one of the mandated steps adopted by the practitioners to analyze and predict the behavior of the system with unforeseen inputs. This has helped develop robust software that is immune against a variety of attacks like buffer-overflow~\cite{bufferOverflow}, divide by zero~\cite{diveByzero}, arithmetic overflow~\cite{arithmeticOverflow}. Hardware security faces vulnerability through the inclusion of malicious logic in the design, commonly referred to as trojan horses, which can elude the functional verification of the design. Activation of this malicious logic is triggered by inputs crafted by attackers. Consequently, security testing of hardware designs becomes exceptionally crucial before progressing to the subsequent stages~\cite{iccd2017_concolic,itc2018_concolic,vlsid2018_atpgModelChecking,date2019_concolic}.~\autoref{table:priorWork} presents some of the notable works on security testing of hardware IPs applicable to cloud FPGAs, along with where our contribution fits in. With high level synthesis becoming the new trend for designing customized hardware accelerators, few works have studied security vulnerabilities in HLS IPs and proposed preliminary countermeasures to detect them~\cite{s3cbench,fuzzSystemC}. 
In prior work like~\cite{fuzzSystemC} that deals with the testing and verification at HLS, terms the trojan inserted in high-level design is called synthesizable hardware Trojan (SHT) as it gets manifested as a malicious backdoor in the hardware design. 
Therefore, the problem of Trojan detection in low level RTL design can be appropriately abstracted as finding SHT in high-level design.
The following sections elaborate on two widely recognized security testing methodologies extensively utilized in the software communities and, more recently, achieving success in the hardware community as well, namely Greybox Fuzzing and Concolic Testing.

\subsubsection{Greybox fuzzing} Fuzz testing is a well known technique in the software domain for detecting bugs\cite{afl-cov}. Greybox fuzzing involves instrumenting the code segments at the point-of-interest, and generating interesting test vectors using evolutionary algorithms. A fitness function is used in order to evaluate the quality of a test-vector. Typically, greybox fuzzing is used to improve branch-pair coverage of the design, therefore the codes are annotated at every basic block. A test-vector is regarded as interesting, if it reaches a previously unexplored basic block, or hits it for a unique number of times. The fuzz engine maintains a history-table for every basic block covered so far, and retains interesting test vectors for further mutation/crossover. The test generation process completes once the user-defined coverage goal is achieved. Popular coverage-guided greybox fuzzing ( CGF ) engines like American fuzzy lop ( AFL) have been able to detect countless hidden vulnerabilities in well-tested softwares~\cite{afl}. 
There exists a plethora of works \cite{safl,driller,verifuzz} that have improved the performance of greybox fuzzing by augmenting various program analysis techniques. Fuzz testing has recently been explored for testing RTL designs \cite{rfuzz, directFuzz, hyperfuzz, copilia}, although not much used to test SystemC designs. In \cite{fuzzSystemC} Le et al has proposed a fuzzing-based framework, AFL-SHT, for test generation to detect hardware trojans in high level SystemC designs.

\subsubsection{Symbolic and Concolic Execution} 
\begin{figure}[t]
\centering
\hspace*{\fill}%
\resizebox{0.25\textwidth}{!}{
\begin{minipage}[c][0.8\width]{
  0.28\textwidth}
  \centering
\begin{lstlisting}[style=customc,label={label:symbeg}]
int foo(int i, int j){
//i and j symbolic variables
  int x = 0, y = 0, z;
  z = 2*j;
  if(z == i){
    if (j < 5){
         x = -2; y = 2;
     }
     else 
      { x = 2; y = 2;}
  }
  else 
   { x = -2; y = -2;}
  return 0;
}

\end{lstlisting}
\end{minipage}
}
\hfill
  {
\begin{minipage}[c][1\width]{
  0.25\textwidth}
  \centering
  \includegraphics[width=1\textwidth]{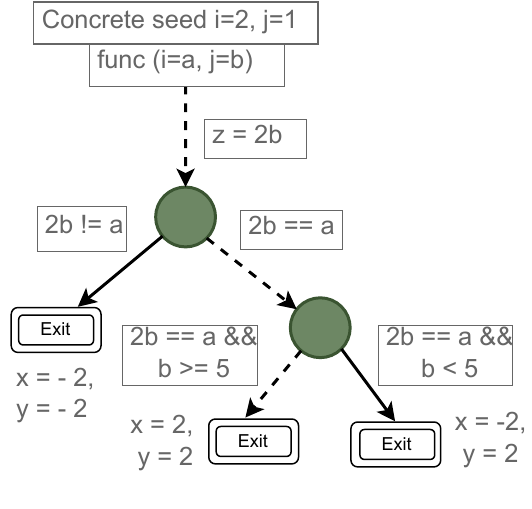}
\end{minipage}}
\hspace*{\fill}%
\caption{(a) Example code snippet. (b) Symbolic and concolic execution flow}
\label{fig:symb-conc}
\end{figure}

Concolic testing~\cite{sen2007concolic} is a scalable way of performing symbolic execution on a program with certain inputs considered concrete, and the rest are symbolic inputs. Symbolic execution in general suffers from scalability issues since the number of path constraints generated, are exponential in terms of the number of conditional statements. In order to avoid costly computations, concolic execution executes the program along the path dictated by concrete input and fork execution at branch points. The path constraints generated in concolic execution have reduced the number of clauses and variables, thereby making it easier for solvers and can penetrate deep into complex program checks. \emph{Driller}~\cite{driller} and \ac{S2E}~\cite{s2e} are examples of engines adopting this approach. We demonstrate concolic execution with a simple example (\autoref{fig:symb-conc}a). A set of program inputs are treated as symbolic variables; the inputs $i$ and $j$ have symbolic values $i=a$ and $j=b$. We choose a random concrete input ($i=2, j=1$) and obtain the execution trace (dotted lines, \autoref{fig:symb-conc}b). The alternative path constraints to be explored symbolically are collected along the execution path guided by concrete inputs, forking the side branches. At each condition, constraints are negated and solved to generate new test cases. The concolic engine terminates after all conditions are covered.
Researchers used concolic testing approaches to detect vulnerabilities in behavioral level RTL designs~\cite{iccd2017_concolic,itc2018_concolic,vlsid2018_atpgModelChecking,date2019_concolic}. Works like~\cite{scalable-rtl, directed_for_rtl} are based on directed test case generation by concolic testing. Concolic test generation for verification of high level designs has only been proposed recently in~\cite{2016_concolic,2018_concolic_sysC,symbolicSystemC}, however it suffers from inherent scalability issues of concolic testing. In~\cite{symbolicSystemC} the authors identify the additional overhead of concolic testing from usage of software libraries, and have restricted the search space within the conditional statements of design. They named their automated prototype as SCT-HTD (SCT-HTD), based on symbolic executer (S2E).

\subsection{Hybrid Test Generation Framework}
Testing a full system is often challenging by individual techniques like fuzz testing and concolic execution when exercising deeper code area to find more bugs in a design. Fuzz testing is efficient at broad coverage using its random mutation techniques but can struggle to uncover intricate corner-case bugs. Concolic execution on other hand has the powerful ability to solve the corner bugs by program analysis and constraint solving. A major impediment to practical concolic execution is speed compared to fuzz testing. The “path explosion” problem resulting from the symbolic loop unrolling is another inherent problem of concolic execution. Thus in recent times, researchers combined these two mainstream testing techniques, integrating the advantages and mitigating the disadvantages of individual techniques. Driller \cite{driller}, Fuzzolic \cite{fuzzolic} are examples of such approaches in the software domain. 
Our recent work, GreyConE~\cite{greycone}, extends this approach to high-level designs (C, C++, SystemC), leveraging the complementary strengths of fuzz testing and concolic execution to enhance test-case generation. 

\subsection{Reference Model Based Testing} 
Reference model-based testing involves using an executable specification, known as a reference model, to predict accurate outcomes based on provided stimuli~\cite{model-based-testing}. The creation of such a model involves various stakeholders, such as the Quality Assurance (QA) Team, Domain Experts, or Testing and Validation Team, depending on project goals.
In MATLAB, a reference model is implemented using a Model block, establishing a model hierarchy with well-defined interfaces for inputs and outputs. This hierarchical structure promotes code reuse, unit testing, parallel builds, and efficient management of large components, minimizing file contention and merge issues.
A reference model typically provides an abstract and partial representation of the desired behavior of a DUT. Test cases derived from this model form an abstract test suite, functioning on the same level of abstraction. However, this suite cannot be directly executed against the DUT. To bridge this gap, an executable test suite must be derived by mapping abstract test cases to concrete ones suitable for direct execution. Some reference model-based testing environments can automatically generate executable test suites directly from the models, enabling effective communication with the system under test. Some recent works utilizes reference model-based testing to secute IoT/cloud data~\cite{mbtaasiot1, mbtiot2, mbtiotprotocols3, mbtsmartcity4}. 

\begin{figure}[ht]
\begin{center}
 \includegraphics[width=.5\textwidth]{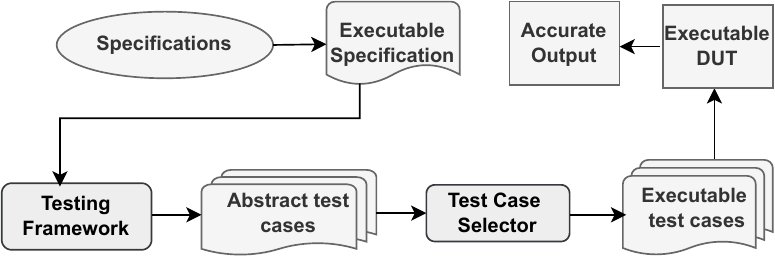}
\end{center}
\caption{Reference Model-based Testing 
(\href{https://en.wikipedia.org/wiki/File:Mbt-process-example.png}{Wikipedia})}
\label{fig:flow_chrt}
\end{figure}

\section{Motivation}
\label{sec:motivation}
Existing research has predominantly focused on testing Hardware HWIPs at RTL or gate-level designs, leaving a significant gap in exploring testing methodologies at higher levels of abstraction, particularly in the context of HLS tailored for FPGA-cloud environments~\cite{farimah2021}. However, vulnerabilities at these higher abstraction levels tend to be more intricate and stealthy, especially when HLS IPs contain malicious functionalities like hidden trojan logic, which can pose substantial risks to cloud environments by leaking sensitive information and tampering with cloud computation and data. Recent investigations into Hardware Trojans within HLS designs highlight attackers' abilities to introduce malicious functionalities that compromise the integrity of the DUT~\cite{s3cbench, fuzzSystemC, symbolicSystemC}.

The background study(~\autoref{table:priorWork}) reveals that existing approaches for testing HWIPs primarily rely on software security testing techniques, such as fuzzing and concolic testing, each with its own limitations~\cite{2016_concolic,2018_concolic_sysC,date2019_concolic,itc2018_concolic,scalable-rtl,symba}. Traditional fuzzing techniques, like AFL, often struggle to satisfy complex conditional checks within a limited timeframe, while concolic engines such as S2E may fail to generate test inputs efficiently due to their slow nature. To address these challenges, our previous work introduced a combined framework named \textit{GreyConE}~\cite{greycone} for efficient test generation in high-level designs. However, \textit{GreyConE} relies on heuristic search methods by both its fuzz engine and concolic engine throughout the entire program space. 

Directed test generation using formal methods holds promise but encounters scalability issues, particularly for large designs due to state space explosion. To address these challenges comprehensively, we propose \textit{GreyConE+}, which combines greybox fuzzing with concolic execution to uncover rare scenarios within complex branches of the design. Prior works on directed test generation for HWIPs have predominantly utilized either fuzz testing or concolic testing alone, with concolic testing facing scalability challenges when targeting rare triggers~\cite{date2019_concolic}. To overcome these hurdles, \textit{GreyConE+} introduces selective instrumentation, which selectively instruments the program on rare targets to enable efficient fuzzing and guide the concolic engine toward specific targets effectively.

Despite the substantial advancements in machine learning (ML)-based trojan detection techniques aiming to reduce reliance on a golden model, we opt for utilizing a reference model for trojan detection. ML-based methods are still in their early stages~\cite{htd-ml, app-htd-ml} and encounter difficulties in automating and ensuring accuracy, which poses risks of misclassifying trojans and allowing them to evade detection. Nonetheless, our approach remains adaptable to situations where a golden model is unavailable, relying on test generation driven by coverage of infrequent triggers to systematically uncover potential hidden malicious circuitry within the design. This motivation underscores the necessity and significance of developing GreyConE+ to address the evolving challenges in securing HLS IPs for FPGA-cloud environments.

\section{Threat Model and Verification Method}
\label{sec:threat-model}
In this section we describe the threat model we consider and the assumed verification methods for IPs in cloud.

\begin{figure}[h]
\begin{center}
\includegraphics[width=.7\textwidth]{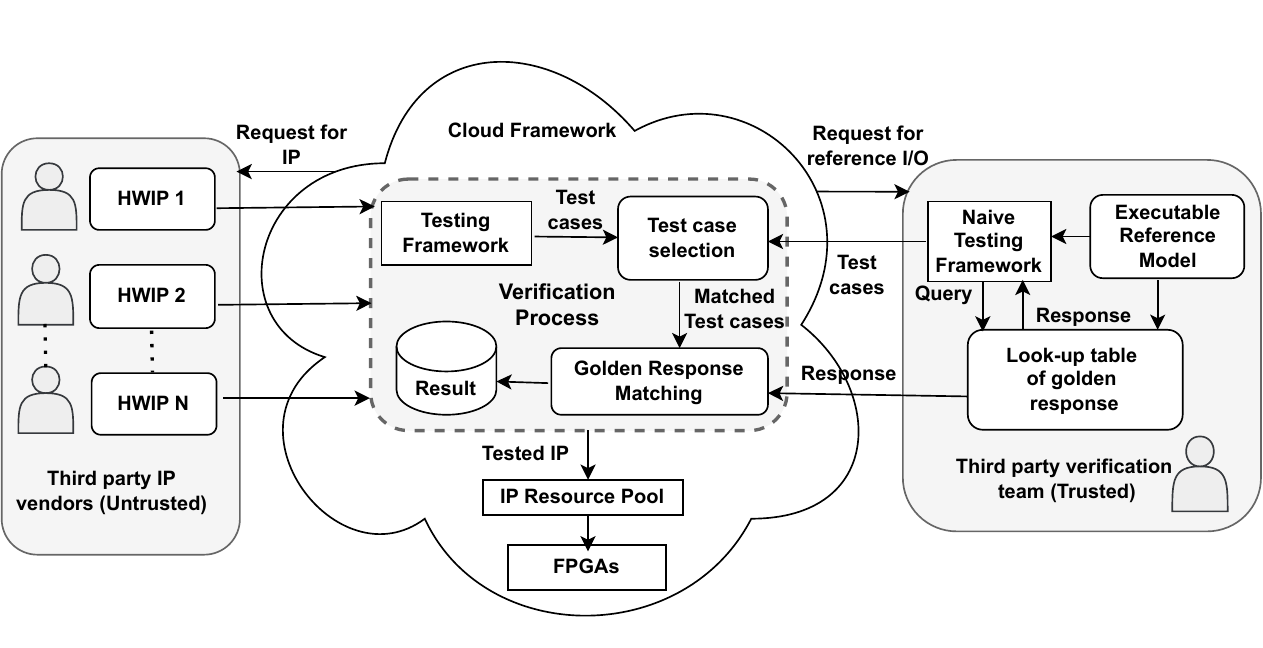}
\end{center}
\caption{Threat Model and Verification Process in Cloud}
\label{fig:threat-model}
\end{figure}

\subsection{Threat Model}

The integration of third-party intellectual property (3PIP) in cloud environments presents significant security challenges. The threat model in~\autoref{fig:threat-model} highlights the risks associated with 3PIPs in the cloud, which may contain hidden vulnerabilities or backdoors due to their external origin. 
The security threats associated with this threat model are similar to those found in untrusted electronics supply chains and Trojan insertions in pre-silicon hardware\cite{state-of-art-HTD,counterfrietIC,trustHW}.
Trojans can be maliciously inserted into third-party IP either intentionally by vendors, who may alter the IP maliciously~\cite{security_of_cloud}, or through inadvertent compromises by external actors who embed Trojans into their products. Vendors might be coerced by criminal organizations or state actors to insert Trojans in exchange for financial rewards or to avoid repercussions. The potential motivations for Trojan insertions include: 1) Stealing confidential information or designs from the cloud IP user, which could be used to develop competing products or services.
2) Disrupting the functionality of the cloud IP, causing outages or crashes that damage the reputation of the company using the IP and potentially lead to financial losses.
3) Gaining unauthorized access to the cloud IP or the system it runs on, which could be exploited for various malicious purposes, such as data theft or installing malware.
4) Slowing down the performance of the IP with delays, making the cloud service provider (CSP) unable to deliver the expected level of service. This can lead to unmet service expectations and significant financial losses.

Therefore, before integrating third-party IP into the cloud FPGA, the cloud service provider (CSP) must conduct thorough security testing to identify any embedded Trojans. This can be achieved using tools and techniques such as static and dynamic analysis to detect malicious code. We propose that the CSP use \textit{ GreyConE+} for this purpose due to its efficient test generation capabilities. However, the CSP currently lacks a definitive standard (golden IP) for testing. To address this, we consider a verification team, acting as a third-party verification team (3PVT) for the CSP, which possesses accurate specifications and a trustworthy golden reference model for the IP. The subsequent section elaborates on the verification process of 3PIPs within the cloud under this threat model.

\subsection{Verification Process in Cloud}
In accordance with our threat model, the verification process in the cloud relies on Reference Model-based Testing. The CSP depends on the 3PVT to assess the integrity of HLS IPs through testing based on a reference model avaialable with the 3PVT. To maintain confidentiality, the 3PVT generates an encrypted look-up table (LUT) from the golden reference model of the IP rather than selling the golden reference IP. The LUT contains golden responses and the design information cannot be extracted from this table. The 3PVT shares the encrypted LUT and a set of abstract test vectors, generated using their testing tool. The CSP utilizes its own testing tool, \textit{GreyConE+}, which produces more streamlined and efficient test cases, which are a subset of those provided by the 3PVT. The CSP matches its executable test cases with those from the 3PVT's test vectors and seeks corresponding golden responses from the LUT, completing the verification process. This independent verification helps detect potential trojans or vulnerabilities, ensuring the integrity of FPGA design IPs and enhancing HWIP security. Importantly, the CSP circumvents the expense of acquiring a costly golden IP, thereby increasing the cost-effectiveness of the verification process while also establishing a secure and well-regulated workflow, as depicted in~\autoref{fig:threat-model}.

\section{GreyConE$+$: Scalable Test Generation Framework}
\label{sec:framework}
We propose \textit{GreyConE+}, a testing framework designed for Third Party Hardware Intellectual Properties (HWIP) implemented in high-level synthesizable languages like SystemC/C++, intended for deployment in cloud-based FPGAs. Building upon our prior framework, \textit{GreyConE}~\cite{greycone}, \textit{GreyConE+} incorporates customized instrumentation steps and combines the strengths of greybox fuzzing and concolic execution. 
\textit{GreyConE+} differs from  \textit{GreyConE} in the following aspects:

\textit{1)}  While \textit{GreyConE} prioritizes overall code coverage, particularly emphasizing branch coverage, \textit{GreyConE+} targets coverage of rare conditions in program, that are potential sites of vulnerabilities,  to enhance coverage of critical areas.

\textit{2)} \textit{GreyConE+} utilizes a selective instrumentation approach during pre-processing, distinguishing it from its predecessor \textit{GreyConE}, which emphasizes on full instrumentation. Specifically, \textit{GreyConE+} instruments the rare scenarios in the program  that are less likely to be encountered during normal program execution while avoiding unnecessary instrumentation in non-critical areas to optimize overall performance.

\textit{3)} \textit{GreyConE+} utilizes stricter switching conditions between fuzz engine and concolic engine, alternating test cases based on threshold time and execution counts, as opposed to \textit{GreyConE} which only considers threshold time for switching. The stringent switching rules of \textit{GreyConE+} facilitate more thorough exploration of the target at each stage.

The comprehensive workflow of GreyConE+ is depicted visually in~\autoref{fig:flow_chrt}, comprising five primary components:
1) Identification of Security Targets, 2) Selective Instrumentation, 3) Greybox Fuzzing, 4) Concolic Execution, and 5) Rare Targets Coverage Checker.
 The fundamental steps in \textit{GreyConE+}'s methodology for activating security targets and detecting trojans are detailed as follows:

\begin{figure}[h]
\begin{center}
\includegraphics[width=1\textwidth]{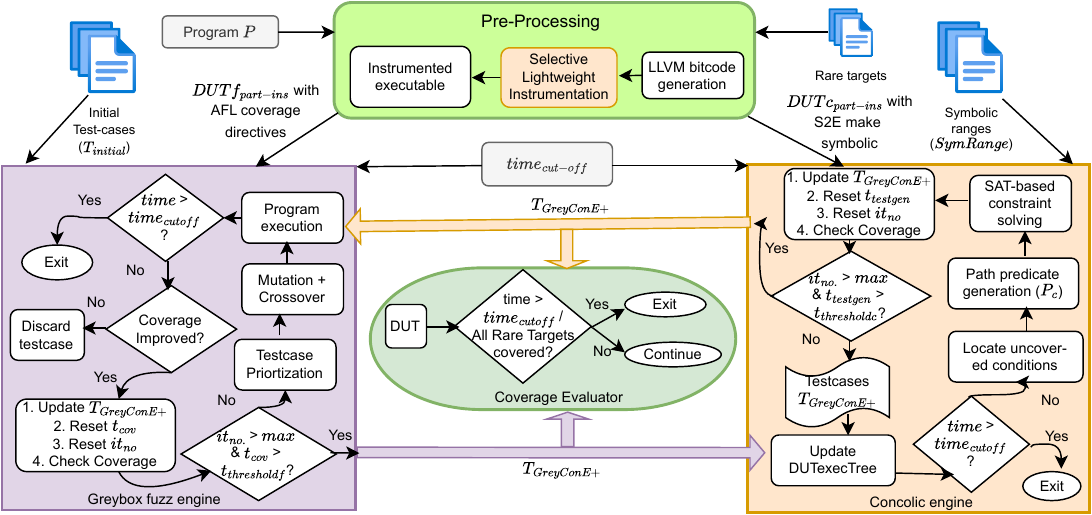}
\end{center}
\caption{\emph{GreyConE+} test generation framework --- Selectively instrumented DUT is generated individually at pre-processing stage and fed to both the test engines.\emph{Fuzz engine} is fed with initial test-cases and \emph{Concolic engine} starts when invoked with fuzzer test-cases. Test-cases are fed back and forth between \emph{Fuzz engine} and \emph{Concolic engine} to accelerate the search space exploration. The \emph{Coverage Evaluator} checks for coverage of the targets whenever \emph{GreyConE+} generates a new test-case.} 
\label{fig:flow_chrt}
\end{figure}

\subsection{Identifying Security Targets from Random Testing}
Traditionally random simulation approach is employed to identify rare branches that may serve as potential sites for
security threats, such as hosting hardware trojans~\cite{directed_for_rtl,date2019_concolic,itc2018_concolic}. 
In this process, random simulation is executed on circuits with rare conditions and infected with trojans supplying random inputs to the designs. 
In the \textit{GreyConE+} framework, random inputs are provided to the HLS designs written in high level languages (SystemC, C++), and random simulation is performed to identify hard-to-reach locations where vulnerabilities may exist. 
The random simulation runs a series of random test cases, activating a broad range of scenarios while leaving rare scenarios—considered security targets—untouched.
The security targets are identified through manual interpretation of cumulative coverage reports (e.g., \texttt{lcov} report), which provide data on any uncovered rare sections of the code. This analysis helps pinpoint specific areas within the design that require instrumentation for further testing to enhance security coverage.

\subsection{Selective Instrumentation for Targeted Testing}

In the context of constructing and testing complex programs, focusing on specific segments or components of the program or hardware description for coverage is advantageous. This approach, known as partial instrumentation, involves selectively instrumenting crucial sections or components while avoiding resource-intensive full design instrumentation~\cite{AFL++}. 
In our framework, we adopt a strategy of partial instrumentation to selectively specify which parts of the design should be instrumented during compilation. This strategy allows to focus on critical areas of interest.
As shown in~\autoref{fig:flow_chrt}, within \textit{GreyConE+} framework, we introduce a methodology to specify rare targets for selective instrumentation during the preprocessing stage.
This approach is facilitated by static analysis tool \texttt{LLVM}~\cite{LLVM}, generating intermediate representation (IR).  
Upon receiving the input targets,\textit{GreyConE+} integrates \textsc{INSTRIM} patches~\cite{instrim} to enhance instrumentation efficiency within \texttt{LLVM}. 
These patches provide an optimized method to select basic blocks for instrumentation of the targeted nodes, leveraging analysis based on Dominator Trees~\cite{dominator-tree} to avoid placing unnecessary instrumentation. This optimization typically reduces the number of instrumented locations by at least half compared to standard instrumentation methods, leading to improved fuzzing performance in terms of speed and effectiveness.
Coverage-guided greybox fuzzers and concolic execution engines like S2E operate on these selectively instrumented executables. By focusing instrumentation on specific program areas, we can guide the fuzzer to prioritize important parts of the codebase. This strategic approach prevents the fuzzer from wasting resources on uninteresting or irrelevant code paths, thereby optimizing the fuzzing process for more effective vulnerability discovery.

After specifying the targets, the basic instrumentation steps are performed at the pre-processing step (\autoref{fig:flow_chrt}).
The \texttt{LLVM} generated \acp{IR} are fed to \texttt{afl-clang-fast}, based on \texttt{clang}~\cite{clang}, a front end compiler for languages like C, C++, SystemC, and others.
\texttt{afl-clang-fast} automatically injects control flow statements onto targeted conditional statements at runtime, generating instrumented executables.
Coverage-guided greybox fuzzers like AFL and concolic execution engines like S2E operate on these instrumented executables. The instrumentation allows test engines to monitor the execution flow triggered by input, determining whether new parts of the program are covered.
Selective instrumentation is critical for maximizing the efficiency and effectiveness of fuzzing and concolic execution. By guiding test inputs towards critical program paths, we enhance the likelihood of identifying vulnerabilities while minimizing unnecessary testing overhead associated with unimportant code segments.

{\small
\begin{algorithm}[!ht]
\footnotesize
\SetAlgoLined
\KwData{Partially instrumented DUT $DUT_f{part-ins}$, User provided test-inputs $T_{initial}$, User defined time bound $time_{cut-off}$} 
\KwResult{{$T_{fuzzed}$}       
   \Comment{Interesting test-inputs queue}}
$T_{fuzzed} \gets T_{initial}$ \Comment{Initialization of the AFL's test-inputs queue}\\ 
\While{time $ \leq time_{cut-off}$}{
    \For{$\tau \in T_{fuzzed}$} 
   {
        \Comment{Mutate $\tau$ to generate test-cases based on the energy parameter}\\
        $K\gets$ CALCULATE\_ENERGY($\tau$)\\
        \For{$i \in \{1,2,\dots,K\}$}{
            $\tau'\gets$ MUTATE-SEED($\tau$) \Comment{$\tau'$ denotes the mutated test case} \\
            \If{IS-INTERESTING($DUT_f{part-ins}, \tau'$)}
            {
            $T_{fuzzed}\gets$ $T_{fuzzed}\cup \tau'$
            \Comment{$\tau'$ is interesting if it improves branch coverage}\\}
        }
        }
    }
    \Return $T_{fuzzed}$
\caption{FUZZER($DUT_f{part-ins}$, $T_{initial}$)}
\label{algo:AFL_FUZZ}
\end{algorithm}
}

\subsection{Greybox Fuzzing on Rare Targets}
We have configured the greybox fuzzer AFL specifically to work with partially instrumented binaries compiled with coverage directives targeting interested program parts.
In Algorithm~\ref{algo:AFL_FUZZ}, we outline the overall flow of Greybox fuzzing by fuzzer like \ac{AFL}. At first, we provide the high-level DUT partially instrumented and compiled with interesting rare targets, $DUT_{f_{part-ins}}$, and a user-provided test-set, $T_{initial}$, to the fuzzing framework. 
The \textit{CALCULATE-ENERGY} function assigns energy to each of the initial test inputs in $T_{initial}$ on the basis of external features of the test case such as the execution time, bitmap coverage, depth within the fuzzing hierarchy. A test case that is fast, covers more branches and has more depth, is given more energy. AFL then decides the number of random fuzzing iterations for that test case. 
\ac{AFL} performs \texttt{deterministic mutations} (bit flipping, byte flipping, arithmetic operations, value substitutions) and \texttt{havoc mutations} (random bit/byte value mutations, sub-sequence cloning or deletion) on the test case $\tau$ to generate new test cases with the help of the \textit{MUTATE-SEED} function.
\ac{AFL} uses branch-pair as a fitness metric to determine the quality of test inputs. 
For each branch-pair, \ac{AFL} maintains a hash-table entry of the number of times it is hit. The \textit{IS-INTERESTING} function evaluates mutated test cases to determine if they cover new branch pairs or exhibit unique behavior compared to past observations. Test inputs deemed \texttt{interesting} (improving branch coverage or exhibiting unique behavior) are retained for further fuzzing.
The fuzzing loop continues until either no more interesting test cases can be found or the user-defined time bound ($time_{cut-off}$) is reached.
\ac{AFL} maintains all interesting test-inputs in the queue $T_{fuzzed}$.

\subsection{Concolic Execution with Rare Targets}
\label{label:s2e_engine}
In our work, we utilize \ac{S2E} as our concolic execution engine to generate test cases. \ac{S2E} consists of a concolic virtual machine based on QEMU~\cite{qemu} and and a symbolic execution engine powered by KLEE~\cite{klee}, enabling seamless transitions between concrete and symbolic execution modes. 
We provide it the high level design {DUT which is partially instrumented and compiled with symbolic rare targets, $DUT_{c{part-ins}}$, a set of test-cases $T_{initial}$ and a symbolic reference file $SymRange$ that specifies which byte ranges of the input file to make symbolic.
The \textit{CONC-EXEC} execute the DUT with all test-cases $T_{initial}$ (for \textit{GreyConE+} $T_{initial}$ is the set of test-cases generated by the fuzzer) to obtain concrete execution traces. 
\ac{S2E} maintains an execution tree $DUT_{execTree}$ and identifies all the true and/or false edges of conditional nodes which are not covered by $T_{initial}$.
\ac{S2E} then assigns symbolic values to those predicates internally and the rare targets. The \textit{COND-PREDICATE} constructs the path constraints for the uncovered edge of a condition, forks a new thread and invokes SAT-solver (\textit{CONSTRAINT-SOLVER}) to generate the test-case. 
\ac{S2E} selects paths for exploration in a depth-first search order, based on its coverage analyzer, aiming to maximize coverage. So, the final test-cases reported by \ac{S2E} ideally should cover all conditions of $DUT_{execTree}$. We outline this approach adapted by concolic engine S2E in Algorithm~\ref{algo:concolic_execution}.

{\small
\begin{algorithm}[t]
\footnotesize
\SetAlgoLined
\KwData{Partially instrumented DUT, $DUT_c{part-ins}$, User provided test-inputs $T_{initial}$, $SymRange$ specifies byte ranges of the input file to make symbolic.}
\KwResult{{$T_{concolic}$} \Comment{Set of test-cases generated by the concolic engine}} 
$DUT_{execTree} \gets \phi$ \Comment{Execution tree for DUT}\\
\For{$\tau \in T_{initial}$}
{
    \Comment{Update DUT's execution tree with path traces obtained from concrete execution of initial test inputs}
    $P_{trace} \gets $ CONC-EXEC($DUT_{c{part-ins}}$,$\tau$)\\
    $DUT_{execTree} \gets DUT_{execTree} \cup P_{trace}$
}
\For{uncovered cond $c \in DUT_{execTree}$}
{
    \Comment{Perform symbolic execution steps targeting uncovered conditional statements}\\
   $p_{c} \gets$ COND-PREDICATE($c$,$SymRange$) \\
   $t_{new}$ $\gets$ CONSTRAINT-SOLVER($p_c$) \Comment{$t_{new}$ is newly generated test-case by the concolic engine}\\
    $T_{concolic}$ $\gets$ $T_{concolic}$ $\cup$ $t_{new}$
    
} 
\Return $T_{concolic}$
\caption{CONCOL-EXEC($DUT_{c{part-ins}}$,$T_{initial}$, $SymRange$)}
\label{algo:concolic_execution}
\end{algorithm}
}

\subsection{Interleaved Fuzzing and Concolic Testing (\textit{GreyConE+})}
GreyConE+ integrates a strategic fusion of greybox fuzzing and concolic execution to effectively address the inherent challenges associated with each technique in isolation (\autoref{fig:flow_chrt}). 
The interleaved approach starts by feeding both the fuzzer and the concolic engine with the selectively instrumented instances of the DUT, $DUT_{f{part-ins}}$ and $DUT_{c{part-ins}}$ respectively. The selective instrumentation enables efficient fuzzing and symbolic execution of specific targets within the DUT.
Concolic engine usually performs depth-first search for search-space exploration. A set of random test cases leads to invoking the SAT/SMT solver frequently for generating test cases to cover unexplored conditions in $DUTexecTree$.
To avoid this slowdown and support concolic execution effectively, we leverage random fuzzing of the DUT to obtain initial seed inputs for the concolic engine (\textit{CONCOL-EXEC}). We initialize our fuzz-engine (FUZZER) with an initial set of test cases ($T_{initial}$).
The fuzzing process is iterated for a predetermined number of cycles, $\#it_{f}$, generating new test cases while adhering to a specified time threshold, $time_{threshold_f}$, for each iteration. The generated fuzzing seeds are prioritized based on their contribution to exploring new basic block coverage.

The concolic engine (\textit{CONCOL-EXEC}) is invoked with the fuzzing-generated seeds which serve as guides for the concolic engine, indicating paths to explore and variables to symbolize for the SAT solver. 
Concolic execution engines like S2E have support for alternating back and forth between symbolic execution and concrete execution which allows the program to run with
concrete seed inputs.  
S2E initially performs a concrete execution using a fuzzing-generated seed and then symbolizes the unvisited rare nodes.
To manage the challenge of path explosion during concolic execution, we limit the number of iterations, $\#it_{c}$, of concolic engine. 
In each iteration, a seed from the fuzzing-generated set is chosen to execute the DUT in concrete mode, directing the concolic engine towards rare paths.
Concolic execution from this state yields new test cases closer to elusive targets, leveraging the concrete execution trace retained from previous iterations.
If no new seeds are available for concrete execution, a new fully symbolic state is forked.
A predefined time threshold ($time_{threshold_c}$) is set for the concolic engine for test generation, and if rare triggers remain unexplored, test cases generated by S2E are reintroduced to the fuzzing engine. This iterative
process persists until the termination condition is satisfied or $time_{cut-off}$ is reached, ensuring thorough exploration of the DUT’s behavior space (\autoref{fig:flow_chrt}).
By interleaving greybox fuzzing and concolic execution in this orchestrated manner, GreyConE+ systematically explores the design space of the DUT, effectively addressing the complexities and challenges associated with each testing technique. This approach ensures comprehensive test coverage and facilitates the detection of elusive vulnerabilities or targets within the DUT.

\begin{algorithm}[h]
\footnotesize
\SetAlgoLined
\KwData{$DUT_{exe-cov}$, $tb$, $T_{GreyConE+}$, $list_{rare-nodes}$}
        $coverage_{info} \gets CovSIMULATOR(DUT_{exe-cov},tb, T_{GreyConE+})$ \\
$count \gets 0$ \\
        \For{$rn \in list_{rare-nodes}$} {
            \If{$covered(coverage_{info},rn)$} {
                $count++$ \Comment{Count of rare nodes covered}\\
                }
                }
        \If{$count \doteq |{list_{rare-nodes}|}$ }{
            All rare targets are coverage\\
            \texttt{break}
        }
$coverage_{GreyConE+} = \texttt{reportCoverage}(coverage_{info})$ \\
\Return ($coverage_{GreyConE+}, count$)
\caption{COV-EVALUATOR ($DUT_{exe-cov}$, $tb$, $list_{rare-nodes}$, $T_{GreyConE+}$)}
\label{algo:coverage}
\end{algorithm}

\subsection{Coverage Evaluation}
The overall approach by \textit{GreyConE+} to check coverage of rare targets as well return overall branch coverage obtained is shown in Algorithm~\ref{algo:coverage}.
The designs are simulated iterating over all the generated \textit{GreyConE+} test cases using a testbench for each design. 
Coverage analyzers like \textit{aﬂ-cov}~\cite{afl-cov} which generate gcov code coverage results for a targeted binary, give the provision to check
coverage of any specifc lines or function of code.
As depicted in Algorithm~\ref{algo:coverage}, the coverage evaluator is fed with the DUT executable compiled with coverage support $DUT_{exe-cov}$, the testbench for the design $tb$ , the list of rare nodes $list_{rare-nodes}$ and the \textit{GreyConE+} generated test-cases.
It generates the coverage information in $coverage_{info}$ of the execution traces obtained by running the design with the generated test-cases. The overall code coverage report is generated from this $coverage_{info}$ by the $reportCoverage$ function. The final coverage report is recorded in $coverage_{GreyConE+}$ and is returned along with the number of rare nodes covered ($count$)

\begin{algorithm}[h]
\footnotesize
\SetAlgoLined
\KwData{$DUT$ is Device Under Test, $RefLUT$ is the reference Look-Up Table, $t$ is the new test-case of \textit{GreyConE+}}
        $out_{ref} \gets RefLUT(t)$ \\
        $out_{DUT} \gets SIMULATOR(DUT,t)$ \\
        \If{$out_{DUT} \neq out_{ref}$ }{
            \Return True\\ } \Comment{Trojan gets detected}\\
        \Else{
            \Return False\\}
\caption{HT-DETECTOR ($DUT, RefLUT, t$)}
\label{algo:htdetector}
\end{algorithm}

\subsection{Trojan Detection}

The test cases generated by \textit{GreyConE+} are used to detect trojans in IPs obtained from third-party vendors in the cloud.
CSP-acquired the golden response (I/O mapping) in the form of a look-up table along with a set of test vectors from the third-party verification team (3PVT). This enables the detection of anomalies resulting from concealed circuitry in rare locations. The optimized \textit{GreyConE+} test cases are a subset of those provided by the 3PVT for input-output mapping. These tailored test cases are then executed on the design to examine the potential presence of trojan payloads in rare locations. Subsequently, a comparison between the outcomes of the \textit{GreyConE+} test cases and those of the golden response is conducted to discern the possible existence of trojans. Infected designs may display disparate outputs, signaling functionality corruption, information leakage, or inconsistencies compared to the golden outputs. The trojan detection mechanism for \textit{GreyConE+} follows Algorithm~\ref{algo:htdetector}.

\section{Experimental setup and design}
\label{sec:results}

\subsection{Experimental setup}

We implement GreyConE+ using state-of-art software testing tools: AFL~\cite{afl} for greybox fuzzing and \ac{S2E}~\cite{s2e} to perform concolic execution. Instrumentation steps are performed using $LLVM 14$ along with $Clang 11$. For robust coverage measurements, we cross-validated our results using a combination of coverage measuring tools: namely \emph{afl-cov-0.6.1}~\cite{afl-cov}, \emph{lcov-1.14}~\cite{Lcov}, and \emph{gcov-9.4.0}~\cite{Gcov}. 
To calculate the line of code (LOC), we use the tool \emph{cloc 1.82}. Experiments are performed on 64-bit linux machine having $i5$ processor and 32 GB RAM clocked at 3.60 GHz.

\subsection{Benchmark characteristics}
We conducted an assessment of \textit{GreyConE+} across a diverse range of designs implemented in SystemC and C++, including \textit{SCBench}~\cite{scbench}, \textit{S2CBench}~\cite{s2cbench}, \textit{S3CBench}~\cite{s3cbench_benchmark}, and Rosetta~\cite{rosetta}, encompassing both trojan-free and trojan-infected designs. These benchmarks include branches (targets) that are hard to cover, presenting a test generation complexity that is reasonable. Our experiments involved a total of $19$ designs, with $15$ originating from the SystemC benchmark and the remaining from the C++ Rosetta benchmark suite. Notably, we present our findings on the extensive SystemC benchmark, as well as on the intricate Rosetta suite benchmark. 
The designs undergo subtle adjustments in their input/output methodology to align seamlessly with the GreyConE framework.
The choosen designs have the basic characteristics of cloud IPs that run on FPGAs and the trojans in the designs covers the threats associated with malicious cloud IPs.
This showcases the effectiveness of our approach. The characteristics of the benchmarks are detailed below. 

\begin{table}[t]
\centering
\caption{Trojan types --- Combinational: Comb., Sequential: Seq.}
\begin{tabular}{cccc}
\toprule
Trojan & Trigger & Payload & Severity\\
\midrule
\ot{cwom}   & Comb. & Comb.   & Low      \\ 
\vl{cwm}    & Comb. & Seq. & High     \\ 
\sep{swom}   & Seq.    & Comb   & Low      \\ 
\ct{swm}    & Seq    & Seq. & High     \\
\bottomrule
\end{tabular}
\label{tab:trojan_type}
\end{table}

\subsubsection{SystemC Designs}

The SystemC designs are carefully chosen from a diverse array of application domains, encompassing various open-source hardware designs. The selected benchmarks exhibit a wide range of characteristics, including \textit{adpcm}, \textit{FFT}, \textit{idct} (image processing cores), \textit{aes}, \textit{md5c} (cryptographic cores), \textit{Quick sort} (data manipulation), \textit{decimation} (filters), and \textit{uart} (communication protocols). These designs span multiple application domains, including security, CPU architecture, network, and DSP, effectively covering the core features and data types of the SystemC language.The trojans embedded in these designs serve different purposes, including Denial-of-Service, Information leakage, and corrupted functionality. These trojans are categorized based on their triggering mechanisms, as outlined in~\autoref{tab:trojan_type}. For trojan types with a memory-based payload, the trojan remains active for an extended period even when the trigger condition is no longer active, and severity levels are defined based on this characteristic. The SystemC designs are synthesizable to RTL using any commercial HLS tool. ~\autoref{tab:bench_char} provides a characterization of the S3C benchmark for the trojan-induced circuit and the three designs from S2C benchmark ($FFT, idct, md5c$) where we have induced trojans with necessary modifications in the designs.  

\begin{table}[t]
    \centering
    
    \caption{HLS synthesized hardware characterization of S3C and S2C benchmarks. Trojan types: \vl{cwm}, \ot{cwom}, \ct{swm} and \sep{swom}}
    \resizebox{0.85\textwidth}{!}{%
    \begin{tabular}{ccccccccc}
        \toprule
        \multirow{2}{*}{\textbf{Benchmark}} &
        \multirow{2}{*}{\textbf{Type}} &
        \multicolumn{3}{c}{\textbf{SystemC characterization}} &
        \multicolumn{4}{c}{\textbf{HLS synthesized hardware}} \\
        \cmidrule(lr){3-5}
        \cmidrule(lr){6-9}
        & & \textbf{Branches} & \textbf{Lines} & \textbf{Functions} & \textbf{LUTs} & \textbf{Registers} & \textbf{Nets} & \textbf{Critical path(ns)}
        \\ 
        \midrule
        \multirow{2}{*} {adpcm} & \ct{swm} & 28 & 316 & 6 & 120 & 118 & 394 & 3.801
        \\ & \sep{swom} & 30 & 317 & 6 & 163 & 240 & 588 & 3.019
        \\
        \cmidrule(lr){1-9}
        aes & \ot{cwom} & 68 & 480 & 13 & 2886 & 4772 & 9039 & 7.668
        \\
        \cmidrule(lr){1-9}
        decimation & \ct{swm} & 94 & 463 & 3 & 3108 & 1741 & 634 & 8.702 \\
         \cmidrule(lr){1-9}
        \multirow{2}{*}{disparity} & \vl{cwm} & 96 & 450 & 4 & 2259 & 1166 & 540 & 3.750
        \\ & \ot{cwom} & 96 & 451 & 4 & 2259 & 1166 & 540 & 3.750 \\
        \cmidrule(lr){1-9}
        FFT\_fixed & \ot{cwom} & 38 & 336 & 1 & 3661 & 2891 & 338 & 7.493 \\
       \cmidrule(lr){1-9}
        filter\_FIR & \ot{cwom} & 16 & 242 & 4 & 89 & 59 & 213 & 7.46 \\
       \cmidrule(lr){1-9}
        idct & \ot{cwom} & 60 & 452 & 4 & 11509 & 2268 & 98 & 6.127 \\
       \cmidrule(lr){1-9}
         \multirow{3}{*}{interpolation} & \ot{cwom} & 30 & 301 & 3 & 1071 & 595 & 212 & 8.331
        \\ & \ct{swm} & 30 & 301 & 3 & 612 & 570 & 212 & 8.321 
        \\ & \sep{swom} & 30 & 302 & 3 & 612 & 569 & 212 & 8.321
        \\
        \cmidrule(lr){1-9}
         \multirow{2}{*} {kasumi} & \ct{swm} & 38 & 414 & 12 & 1385 & 958 & 272 & 8.016 \\
         & \ot{cwom} & 38 & 414 & 12 & 1431 & 987 & 273 & 9.266 \\
        \cmidrule(lr){1-9}
        md5c & \ot{cwom} & 34 & 490 & 6 & 84273 & 25476 & 614 & 8.750 \\
       \cmidrule(lr){1-9}
         \multirow{3}{*} {sobel} & \vl{cwm} & 46 & 391 & 2 & 801 & 390 & 112 & 3.750
        \\ & \ot{cwom} & 46 & 289 & 2 & 801 & 390 & 112 & 3.750
        \\ & \ct{swm} & 36 & 391 & 2 & 484 & 285 & 112 & 2.189 
        \\
        \cmidrule(lr){1-9}
         \multirow{2}{*} {uart} & \ct{swm-1} & 48 & 328 & 3 & 549 & 196 & 1336 & 2.766 
         \\ & \ct{swm-2} & 50 & 328 & 3 & 566 & 190 & 1336 & 4.367 \\
    \bottomrule
    \end{tabular}
    \label{tab:bench_char}
    }
\end{table}
\subsubsection{Rosetta HLS Benchmark}
The Rosetta~\cite{rosetta} benchmark stands as a practical High-Level Synthesis Benchmark Suite tailored for Software Programmable FPGAs. Within Rosetta, designs represent fully-developed applications, complete with realistic performance constraints and optimization leveraging the advanced features of contemporary HLS tools. Presently, Rosetta comprises six authentic benchmarks drawn from machine learning and video processing domains, showcasing the competitive energy efficiency of FPGAs in comparison to CPUs and GPUs. The benchmark suite is implemented in C++, and~\autoref{tab:rosetta_char} provides a comprehensive overview of four designs of the current Rosetta collection along with their HLS synthesized characteristics using commertial HLS tool. The Rosetta benchmark enables the running of designs on FPGAs through its proprietary synthesis methods.

\begin{table}[]
\caption{The C++ characteristics and the HLS synthesized characteristics of the HLS benchmark suite Rosetta\cite{rosetta}}
\centering
\resizebox{0.85\textwidth}{!}{%
\begin{tabular}{ccccccccc}
\toprule
\multirow{2}{*}{\textbf{Benchmark}} &
  \multicolumn{3}{c}{\textbf{C++ Characteristics}} &
  \multicolumn{5}{c}{\textbf{HLS Synthesized Characteristics}} \\ \cmidrule(l){2-4} \cmidrule(l){5-9} 
 &
  \multicolumn{1}{c}{\textbf{LOC}} &
  \multicolumn{1}{c}{\textbf{Functions}} &
  \textbf{Branches} &
  \multicolumn{1}{c}{\textbf{\#LUTs}} &
  \multicolumn{1}{c}{\textbf{\#FFs}} &
  \multicolumn{1}{c}{\textbf{\#BRAMs}} &
  \multicolumn{1}{c}{\textbf{\#DSPs}} &
  \textbf{Timing (ns)} \\ \midrule
3D Rendering &
  \multicolumn{1}{c}{452} &
  \multicolumn{1}{c}{17} &
  100 &
  \multicolumn{1}{c}{2496} &
  \multicolumn{1}{c}{1350} &
  \multicolumn{1}{c}{39} &
  \multicolumn{1}{c}{3} &
  8.697 \\ \midrule
Digit Recognition &
  \multicolumn{1}{c}{317} &
  \multicolumn{1}{c}{10} &
  64 &
  \multicolumn{1}{c}{1753} &
  \multicolumn{1}{c}{1355} &
  \multicolumn{1}{c}{0} &
  \multicolumn{1}{c}{0} &
  8.730 \\ \midrule
Face Detection &
  \multicolumn{1}{c}{893} &
  \multicolumn{1}{c}{20} &
  149 &
  \multicolumn{1}{c}{16955} &
  \multicolumn{1}{c}{8793} &
  \multicolumn{1}{c}{622} &
  \multicolumn{1}{c}{35} &
  9.630 \\ \midrule
Spam Filtering &
  \multicolumn{1}{c}{429} &
  \multicolumn{1}{c}{15} &
  125 &
  \multicolumn{1}{c}{4737} &
  \multicolumn{1}{c}{2335} &
  \multicolumn{1}{c}{2} &
  \multicolumn{1}{c}{17} &
  8.670 \\ \bottomrule
\end{tabular}
\label{tab:rosetta_char}
}
\end{table}

\subsection{Design of experiments}

We conduct diverse sets of experiments to evaluate \textit{GreyConE+} and compare its results with standardized baseline techniques, namely the fuzz-testing-based approach AFL, symbolic model checking with S2E, and our previous work, \textit{GreyConE}. The evaluation focuses on two key aspects: 1) Rare targets coverage, and 2) Trojan detection, considering both improved coverage and time speed-up.
We now describe the setup of each experiments:

\textbf{Random Simulation}: 

We conduct random simulations with a series of 10-20 random test cases, considering the design size, on both trojan-free and trojan-inserted benchmarks to identify the rare triggers in the designs.
The automation is wrapped around  \textit{gcov} \cite{Gcov} with the random test cases to fetch data for the overall code coverage achieved in the form of \textit{lcov} \cite{Lcov} HTML report. Manual interpretation of cumulative lcov report provides data on any uncovered rare events.

For the benchmark designs, the rare targets encompass the trojan branches (mostly 1-2 in number) as well as the targets that are deep enough inside the control flow. Designs from SystemC benchmarks like \textit{decimation}, \textit{disparity}, \textit{fft-fixed}, \textit{idct}, contains uncovered branches that are validated independently to be unreachable code segments of the original designs. Designs like \textit{adpcm}, \textit{aes} contains only the trojan branches as rare targets.
In the \textit{Digit Recognition} design of Rosetta, there is a single function not addressed by random simulation, which we identify as the rare scenario for the design. 

\textbf{Baseline 1~(Fuzz testing)}: \textit{AFL} is executed on the SystemC and C++ designs with default algorithmic settings. The initial seed inputs are generated randomly.

\textbf{Baseline 2~(Symbolic execution)}: Similar to Baseline 1, \textit{S2E} is executed on the SystemC and C++ designs with default configurations. Randomly generated seed inputs serve as inputs to \textit{S2E}.

\textbf{GreyConE}: \textit{GreyConE} is executed on the SystemC and C++ benchmarks following the \textit{GreyConE} architecture. The process begins with randomly generated input test cases. The threshold time for the fuzz engine is set to $time_{thresholdf}=5$ seconds, and for concolic engines, it is set to $time_{thresholdc}=10$ seconds, excluding the execution time needed to generate the first seed by each test engine. These parameters are user-defined and configurable.

\textbf{GreyConE+}: \textit{GreyConE+} is executed on the SystemC and C++ benchmarks based on the \textit{GreyConE+} architecture, specifically designed for rare events. Selective instrumentation of the design is performed on the rare targets given as inputs. The iteration limit for each engine is set as $n = 5$, and the threshold time for the fuzzer and the concolic engine is set at $5$ and $10$ seconds respectively, excluding the execution time needed to generate the first seed by each test engine.

The primary goal is to examine how well \textit{GreyConE+} covers rare statements in the design, thereby boosting defenders' confidence in identifying any anomalous behavior, even in the absence of a reference model. We illustrate the outcomes of our experiments through case studies, showcasing the efficacy of \textit{GreyConE+} across different benchmarks.
To ensure a fair comparison with previous state-of-the-art approaches \cite{symbolicSystemC,fuzzSystemC}, we evaluate Trojan detection results as reported in their published works and set the run-time limit as two hours as used by them. The work of Veerana et.al~\cite{s3cbench} also focuses on HLS IPs of S3CBench, though it doesn't directly compare with our technique as they primarily engage in property-based testing. Additionally, we conduct a comprehensive comparison between \textit{GreyConE+} and \textit{GreyConE}, considering both code coverage and Trojan detection capabilities.

\begin{table*}[!tb]
  
    \centering
\caption{Comparison of Rare Targets Coverage of GreyConE+ with baseline techniques AFL\cite{afl} and S2E\cite{s2e} and with GreyConE\cite{greycone} on trojan infected S3CBench\cite{s3cbench_benchmark} and trojan inserted designs from S2CBench\cite{s2cbench}. Trojan types: \vl{cwm}, \ot{cwom}, \ct{swm} and \sep{swom}.}
\resizebox{\columnwidth}{!}{%
\begin{tabular}{cccccccccccc}
\toprule
\toprule
\multirow{2}{*}{\textbf{Benchmark}} &
  \multirow{2}{*}{\textbf{\begin{tabular}[c]{@{}c@{}}\#Rare\\ Triggers\end{tabular}}} &
  \multicolumn{2}{c}{\textbf{AFL}} &
  \multicolumn{2}{c}{\textbf{S2E}} &
  \multicolumn{3}{c}{\textbf{GreyConE}} &
  \multicolumn{3}{c}{\textbf{GreyConE+}} 
 \\ \cmidrule(lr){3-4} \cmidrule(lr){5-6}  \cmidrule(lr){7-9} \cmidrule(lr){10-12} & 
   &
   \textbf{\#cov.} & \textbf{Time(s)} &
   \textbf{\#cov.} & \textbf{Time(s)} &
    \textbf{\#cov.} & \textbf{Time(s)} & \textbf{Phases} &
   \textbf{\#cov.} & \textbf{Time(s)} & \textbf{Phases}
  \\ \cline{1-12}
 \ \ct{adpcm-swm} &
  1 &
  \rt{0} & \rt{TO} & 
  \rt{0} & \rt{TO} & 
  1 & 438 & $fuzz_1$-$conc_1$-$fuzz_2$ &
  1 & \textbf{258} & $fuzz_1$-$conc_1$ \\ 
  \sep{adpcm-swom} & 1 &
  1 & 126.9 & 
  \rt{0} & \rt{TO} & 
  1 & 52 & $fuzz_1$-$conc_1$ & 
  1 & \textbf{20} & $fuzz_1$-$conc_1$
   \\ \midrule
 \ot{aes-cwom} &
  2 &
  \rt{0} & \rt{TO} & 
   \rt{0} & \rt{TO} & 
  2 & 159.07 & $fuzz_1$-$conc_1$-$fuzz_2$ &
  2 & \textbf{32} & $fuzz_1$-$conc_1$
   \\ \midrule
\ct{decimation-swm} &
  4 &
  1 & 3196.5 & 
  1 & 1643 & 
  1 & 839 & $fuzz_1$-$conc_1$ &
  1 & \textbf{132} & $fuzz_1$-$conc_1$
   \\ \midrule
\vl{disparity-cwm} &
  5 &
  \rt{3} & \rt{3426.14} & 
  \rt{0} & \rt{TO} & 
  5 & 3521.60 & $fuzz_1$....$conc2_2$ &
  5 & \textbf{2686} & $fuzz_1$....$conc2_2$ 
   \\ 
  \ot{disparity-cwom} & 5 &
 5 & 3692.2 & 
  \rt{0} & \rt{TO} & 
  5 & 3594.02 & $fuzz_1$....$conc2_2$ &
  5 & \textbf{2319} & $fuzz_1$-$conc_1$-$fuzz_2$ 
  \\ \midrule
\ot{FFT\_fixed-cwom} &
  2 &
  2 & 1385.68 & 
  2 & 2631 &
  2 & 227.47 & $fuzz_1$-$conc_1$-$fuzz_2$ &
  2 & \textbf{161} & $fuzz_1$-$conc_1$ 
   \\ \midrule
\ot{filter\_FIR-cwom} &
  2 &
  1 & 27.8 & 
  1 & 144 & 
  1 & 24 & $fuzz_1$-$conc_1$ &
  1 & \textbf{6} & $fuzz_1$ 
   \\ \midrule
\ot{idct-cwom} &
  7 &
  7 & 36.11 & 
  7 & 422 & 
  7 & 138.2 & $fuzz_1$-$conc_1$ &
  7 & \textbf{14} & $fuzz_1$ 
   \\ \midrule
\ot{interpolation-cwom} &
  10 &
  \rt{0} & \rt{TO} & 
  \rt{0} & \rt{TO} &
  10 & 3657 & $fuzz_1$-$conc_1$-$fuzz_2$ &
  10 & \textbf{2311} & $fuzz_1$-$conc_1$ 
   \\ 
  \ct{interpolation-swm} & 1 &
  1 & 10.2 & 
  1 & 295 & 
  1 & 10.2 & $fuzz_1$ &
  1 & 19 & $fuzz_1$ 
   \\ 
  \sep{interpolation-swom} & 1 &
  1 & 10.2 & 
  1 & 216 & 
  1 & 10.2 & $fuzz_1$ &
  1 & 13 & $fuzz_1$ 
   \\ \midrule
 \ot{kasumi-swm} &
  1 &
  1 & 58.48 & 
  1 & 233 & 
  1 & 58.48 & $fuzz_1$ &
  1 & \textbf{55} & $fuzz_1$
   \\ 
 \ct{kasumi-cwom} & 1 &
   1 & 58.55 & 
  1 & 278 & 
  1 & 58.55 & $fuzz_1$ &
  1 & \textbf{57} & $fuzz_1$ 
   \\ \midrule
\ot{md5c-cwom} &
  4 &
  \rt{0} & \rt{TO} & 
  \rt{0} & \rt{TO} & 
  4 & 577 & $fuzz_1$-$conc_1$ &
  4 & \textbf{574} & $fuzz_1$-$conc_1$ 
   \\ \midrule
 \vl{sobel-cwm} &
  2 &
   \rt{0} & \rt{TO} &  
   2 & 4892 & 
   2 & 2537 & $fuzz_1$....$conc2_2$ &  
   2 & \textbf{682} &  $fuzz_1$....$conc2_2$ 
   \\ 
 \ot{sobel-cwom} & 2 &
   2 & 685.9 & 
  2 & 592 & 
  2 & 612.64 & $fuzz_1$-$conc_1$ & 
   2 & 771 & $fuzz_1$-$conc_1$ 
   \\ 
 \ct{sobel-swm} & 2 &
  2 & 676.8 & 
  2 & 605 & 
  2 & 489.72 & $fuzz_1$-$conc_1$ &
  2 & \textbf{469} & $fuzz_1$-$conc_1$ 
   \\ \midrule
 \ct{uart-swm-1} &
  5 &
  5 & 2963.6 & 
  4 & 145 & 
  5 & 268 & $fuzz_1$-$conc_1$ &
  5 & \textbf{68} & $fuzz_1$-$conc_1$ 
   \\ 
  \ct{uart-swm-2} & 4 &
   4 & 2385.5 & 
  2 & 106 & 
  4 & 274 & $fuzz_1$-$conc_1$ &
   4 & \textbf{119} & $fuzz_1$-$conc_1$ 
  \\ \bottomrule
  \bottomrule
\multicolumn{12}{l}{Note: Column \#cov. denotes the number of rare targets covered} \\
\end{tabular}
\label{tab:cov-comp}
}
\end{table*}

\section{Empirical analysis}
\label{analysis}

\subsection{Covering rare triggers}
In this section, we present the rare target coverage outcomes of \textit{GreyConE+} compared to existing test generation techniques, specifically \textit{AFL}, \textit{S2E}, and \textit{GreyConE}. While the goal of the existing techniques is to encompass all branches, to compare with \textit{GreyConE+}, we terminate these engines once they successfully cover our specified targets or reach a time-out. The chosen benchmarks for evaluation include designs with rare targets that are hard to activate and have reasonable test generation complexity. The targets are selected through random testing and include the rare branches with Trojan areas as well. \textit{GreyConE+} employs both fuzz and concolic engines interchangeably to address rare nodes specified within the partially instrumented design. Similarly, \textit{GreyConE} also utilizes both the fuzzer and the concolic engine in an interleaved manner. However, \textit{GreyConE+} selectively instruments only the specific parts of the code that are infrequently triggered. Consequently, the coverage by \textit{GreyConE+} requires less time compared to \textit{GreyConE} and other individual techniques \textit{AFL} and \textit{S2E}. 

The coverage results on the trojan inserted designs of \textit{S3CBench} and \textit{S2CBench} by each engine are detailed in~\autoref{tab:cov-comp}. The first column of~\autoref{tab:cov-comp} represents the benchmark's name, while the second column shows the number of rare targets for each design including hard-to-reach trojan points, excluding the unreachable sections of the codes present in some of the designs like \textit{FFT\_fixed}, \textit{disparity}, \textit{uart} etc. Columns 3 to 12 showcase the performance of covering the rare triggers by each approach.  Columnn 3,5,7 and 10 represents the maximum number of rare triggers covered and columns 4,6,8 and 11 represents the time required by each approach. The time required by GreyConE+ also includes the duration of the random simulations. 
We mark TO if a technique fails to trigger a single rare node within the time limit. For example, AFL covers 3 rare nodes in $3426.14$ seconds for \textit{disparity-cwom}, while \textit{GreyConE} and \textit{GreyConE+} cover 5 rare nodes in $3521.6$ seconds and $2569$ seconds, respectively and S2E covers none. In this case we mark TO for S2E only.

\textit{GreyConE} and \textit{GreyConE+} alternate between fuzzing ( $fuzz_{n}$) and concolic execution ($conc_{n}$) denoted by phase IDs $n$.
A rare trigger in phase $fuzz_{3}$ implies test generations of  $fuzz_{1}$-$conc_{1}$-$fuzz_{2}$-$conc_{2}$-$fuzz_{3}$ before reaching the target.
 Columns 9 and 12 compares the phases required by \textit{GreyConE} and \textit{GreyConE+} for rare target coverage. 
Hybrid frameworks GreyConE and GreyConE+ sometimes underperforms due to phase switching, Eg: AFL outperforming \textit{GreyConE} in \textit{idct}, and S2E being faster than \textit{GreyConE} in \textit{uart-swm1} and \textit{uart-swm2}.
\textit{ GreyConE+} excels from \textit{GreyConE} due to it's selective instrumentation approach which targets rare triggers and covers rare nodes more efficiently. As observed from the table, even for design like \textit{kasumi}, where both methods complete in $fuzz_1$, \textit{GreyConE+} is faster than \textit{GreyConE}. 
However, as \textit{GreyConE+} requires additional time for random simulation, we observe that for some designs like \textit{interpolation(swm,swom)} and \textit{sobel-cwom}, the time requirement by AFL and \textit{GreyConE} is less than \textit{GreyConE+} to achieve maximum rare trigger coverage.
As evident from~\autoref{tab:cov-comp} \textit{GreyConE+} traverses fewer phases for designs like \textit{adpcm-swm} and \textit{interpolation-cwom} due to its selective instrumentation, which speeds up rare node exploration at each phase, and its modified phase switch condition, which allows for sufficient exploration at each phase. \textit{GreyConE} on the other hand focuses on overall code-coverage and have frequent phase switches.

Similarly~\autoref{tab:cov-compRosetta} and~\autoref{tab:cov-comp-scbench} shows the coverage results for the trojan free designs of \textit{Rosetta} and \textit{SCBench}, which have been customized to be executed by our framework. These tables report the number of testcases generated (\#tc) and the rare branch coverage percentage (Brcov.\%) for the designs.
Rosetta designs are larger in size compared to the \textit{S3CBench}, containing more rare branches but exhibiting less stealthiness than the trojan benchmarks of \textit{S3CBench}. 
~\autoref{tab:cov-comp-scbench} highlights the effectiveness of \textit{GreyConE+} on the large designs of \textit{SCBench} such as \textit{DES} and \textit{RISC-CPU}, which consist of approximately two thousand lines of code. We observe that for \textit{DES}, \textit{AFL} and \textit{S2E} managed to cover 6 out of 9 rare nodes before getting stuck whereas both \textit{GreyConE} and \textit{GreyConE+} successfully covered all the 9 nodes, with GreyConE+ achieving this feat more rapidly.

\begin{table}[]
\centering
\caption{Comparison of Rare Targets Coverage of GreyConE+ with baseline techniques AFL\cite{afl} and S2E\cite{s2e} and with GreyConE\cite{greycone} on HLS bencmark Suite Rosetta\cite{rosetta}}
\resizebox{\columnwidth}{!}{%
\begin{tabular}{|c|c|c|c|c|c|c|c|c|c|c|c|c|c|c|c|c|c|}
\hline
\multirow{2}{*}{\textbf{Benchmark}} &
  \multirow{2}{*}{\textbf{\begin{tabular}[c]{@{}c@{}}Rare\\ Targets\end{tabular}}} &
  \multicolumn{4}{c|}{\textbf{AFL}} &
  \multicolumn{4}{c|}{\textbf{S2E}} &
  \multicolumn{4}{c|}{\textbf{GreyConE}} &
  \multicolumn{4}{c|}{\textbf{GreyConE+}} \\ \cline{3-18}
 &
   &
   \textbf{\#cov.} & \textbf{Time(s)} & \textbf{\#tc} & \textbf{Brcov.(\%)} &
   \textbf{\#cov.} & \textbf{Time(s)} & \textbf{\#tc} & \textbf{Brcov.(\%)} &
   \textbf{\#cov.} & \textbf{Time(s)} & \textbf{\#tc} & \textbf{Brcov.(\%)} &
   \textbf{\#cov.} & \textbf{Time(s)} & \textbf{\#tc} & \textbf{Brcov.(\%)} 
   \\ \hline
3D Rendering & 41 &
  \multicolumn{1}{c|}{8} & 112.56 & 24 & 31.7 
   &
  \multicolumn{1}{c|}{8} & 2275 & 103 & 31.7
   &
  \multicolumn{1}{c|}{18} & 39 & 18 & 45.2
   &
  \multicolumn{1}{c|}{18} & \textbf{12} & 18 & 45.2
   \\ \hline
Digit Recognition &
  1 &
  \multicolumn{1}{c|}{1} & 569 & 24 & 100
   &
  \multicolumn{1}{c|}{1} & 2561 & 66 & 100
   &
  \multicolumn{1}{c|}{1} & 554 & 8 & 100
   &
  \multicolumn{1}{c|}{1} & 579 & 8 & 100
   
   \\ \hline
Face Detection &
  30 &
  \multicolumn{1}{c|}{8} & 246.86 & 16 & 60.7
   &
  \multicolumn{1}{c|}{8} & 2248 & 183 & 60.7
   &
  \multicolumn{1}{c|}{23} & 504 & 73 & 80.6
   &
  \multicolumn{1}{c|}{23} & \textbf{171} & 49 & 80.6
   \\ \hline
Spam-Filtering &
  6 &
  \multicolumn{1}{c|}{6} & 861.12 & 34 & 100
   &
  \multicolumn{1}{c|}{6} & 938 & 6 & 100
   &
  \multicolumn{1}{c|}{6} & 276 & 25 & 100
   &
  \multicolumn{1}{c|}{6} & \textbf{185} & 20 & 100
   \\ \hline
\multicolumn{12}{l}{Note: Column \#cov. denotes the number of rare targets covered} \\
\end{tabular}
\label{tab:cov-compRosetta}
}
\end{table}

\begin{table}[]
\caption{Comparison of Rare Targets Coverage of GreyConE+ with baseline techniques AFL\cite{afl} and S2E\cite{s2e} and with GreyConE\cite{greycone} on large designs from SCBench\cite{scbench} to show the scalability of GreyConE+}
\resizebox{\columnwidth}{!}{%
\begin{tabular}{|c|c|c|c|c|c|c|c|c|c|c|c|c|c|c|c|c|c|c|}
\hline
\multirow{2}{*}{\textbf{Benchmark}} &
  \multirow{2}{*}{\textbf{\begin{tabular}[c]{@{}c@{}}Rare\\ Targets\end{tabular}}} &
  \multirow{2}{*}{\textbf{LOC}} &
  \multicolumn{4}{c|}{\textbf{AFL}} &
  \multicolumn{4}{c|}{\textbf{S2E}} &
  \multicolumn{4}{c|}{\textbf{GreyConE}} &
  \multicolumn{4}{c|}{\textbf{GreyConE+}} \\ \cline{4-19} 
   &  &  &
  \textbf{\#cov.} & \textbf{Time(s)} & \textbf{\#tc} & \textbf{Brcov.(\%)} &
   \textbf{\#cov.} & \textbf{Time(s)} & \textbf{\#tc} & \textbf{Brcov.(\%)} &
   \textbf{\#cov.} & \textbf{Time(s)} & \textbf{\#tc} & \textbf{Brcov.(\%)} &
   \textbf{\#cov.} & \textbf{Time(s)} & \textbf{\#tc} & \textbf{Brcov.(\%)}  \\ \hline
DES & 9 & 2401 &
  \multicolumn{1}{c|}{6} & 28.9 & 291 & 93.0
   &
  \multicolumn{1}{c|}{6} & 11 & 76 & 93.0
   &
   \multicolumn{1}{c|}{9} & 84.86 & 278 & 100
   &  
  \multicolumn{1}{c|}{9} & \textbf{61} & 204 & 100
   \\ \hline
RISC-CPU & 165 & 2056 &
  \multicolumn{1}{c|}{50} & 1088 & 10 & 67.1
   &
  \multicolumn{1}{c|}{68} & 2090 & 272 & 77.3 
   &
   \multicolumn{1}{c|}{74} & 660 & 125 & 81.3
   &
  \multicolumn{1}{c|}{\textbf{77}} & 864 & 43 & \textbf{81.9}
   \\ \hline
SNOW3G &  2 & 522 &
  \multicolumn{1}{c|}{2} & 1586 & 12 & 100
   &
  \multicolumn{1}{c|}{2} & 1728 & 11 & 100
   &
   \multicolumn{1}{c|}{2} & 498 & 8 & 100
   &
  \multicolumn{1}{c|}{2} & \textbf{411} & 6 & 100
   \\ \hline
\multicolumn{12}{l}{Note: Column \#cov. denotes the number of rare targets covered} \\
\end{tabular}
\label{tab:cov-comp-scbench}
}
\end{table}

\noindent
\textbf{Run-Time Speed Up:}
\autoref{fig:greyp-time} illustrates the run-time speedup of \textit{GreyConE+} over \textit{GreyConE}, \textit{AFL}, and \textit{S2E} across benchmark designs where each technique achieves an equivalent number of rare targets coverage within the designated time limit. Our analysis reveals that \textit{GreyConE+} establishes a lower bound on the time required to cover these rare targets compared to \textit{GreyConE}, \textit{AFL}, and \textit{S2E}. 
The speedup achieved by \textit{GreyConE+} aligns with our design methodology:1)\textit{ GreyConE+} prioritizes rare target coverage by the fuzzer, swiftly identifying targets where the fuzz engine stalls and invoking the concolic engine to resolve complex conditions. 2) \textit{GreyConE+} circumvents costly path exploration via concolic execution by leveraging fuzzer-generated seeds, resulting in faster exploration and test-case generation.
Our evaluation employs CPU wall-time metrics for measuring execution duration.

\begin{figure}
    \centering
    \resizebox{\textwidth}{!}{
    \includegraphics{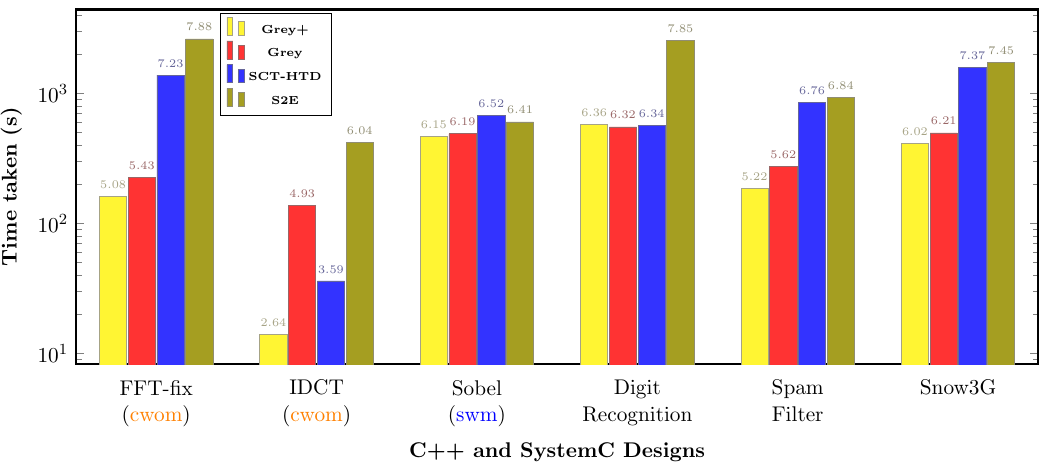}}
        \caption{Time comparison of \textit{AFL}, \textit{S2E}, \textit{GreyConE} and \textit{GreyConE+} for the benchmarks where they cover the same number of rare targets}
\label{fig:greyp-time}
\end{figure}

\noindent
\textbf{Coverage Speed Up:} We presented the results of \textit{GreyConE+} in~\autoref{tab:cov-comp}, showcasing its ability to cover the maximum number of rare targets within a 2-hour time limit, comparing its performance against other test generation techniques.\textit{ GreyConE+} demonstrates superior efficiency by covering a higher number of rare targets in a shorter timeframe compared to other methods.
For instance, in designs such as \textit{adpcm-swm} and \textit{aes-cwom}, \textit{AFL} and \textit{S2E} were unable to cover a single rare node. In \textit{disparity-cwm}, AFL covered only 3 out of 7 rare nodes, while \textit{S2E} failed to cover any. Both \textit{GreyConE} and \textit{GreyConE+} successfully covered the rare node in \textit{adpcm-swm} and \textit{aes-cwom}, as well as 5 rare nodes in disparity. Notably, \textit{GreyConE+} achieved these results in less time than \textit{GreyConE}.
Given that these rare targets often represent uncommon branches, we evaluated each technique's branch coverage throughout the entire 2-hour duration or until all rare targets were covered. Higher percentage coverage indicates higher probability of test-cases detecting bugs hidden in deeper program segments. 
 \autoref{fig:coverageOfGreyConE} presents a comprehensive analysis of branch coverage over this 2 hours timeframe, highlighting \textit{GreyConE+}'s ability to achieve maximum coverage of rare branches in a shorter duration compared to alternative techniques. 

\begin{figure}
    \centering
    \resizebox{\textwidth}{!}{
    \includegraphics{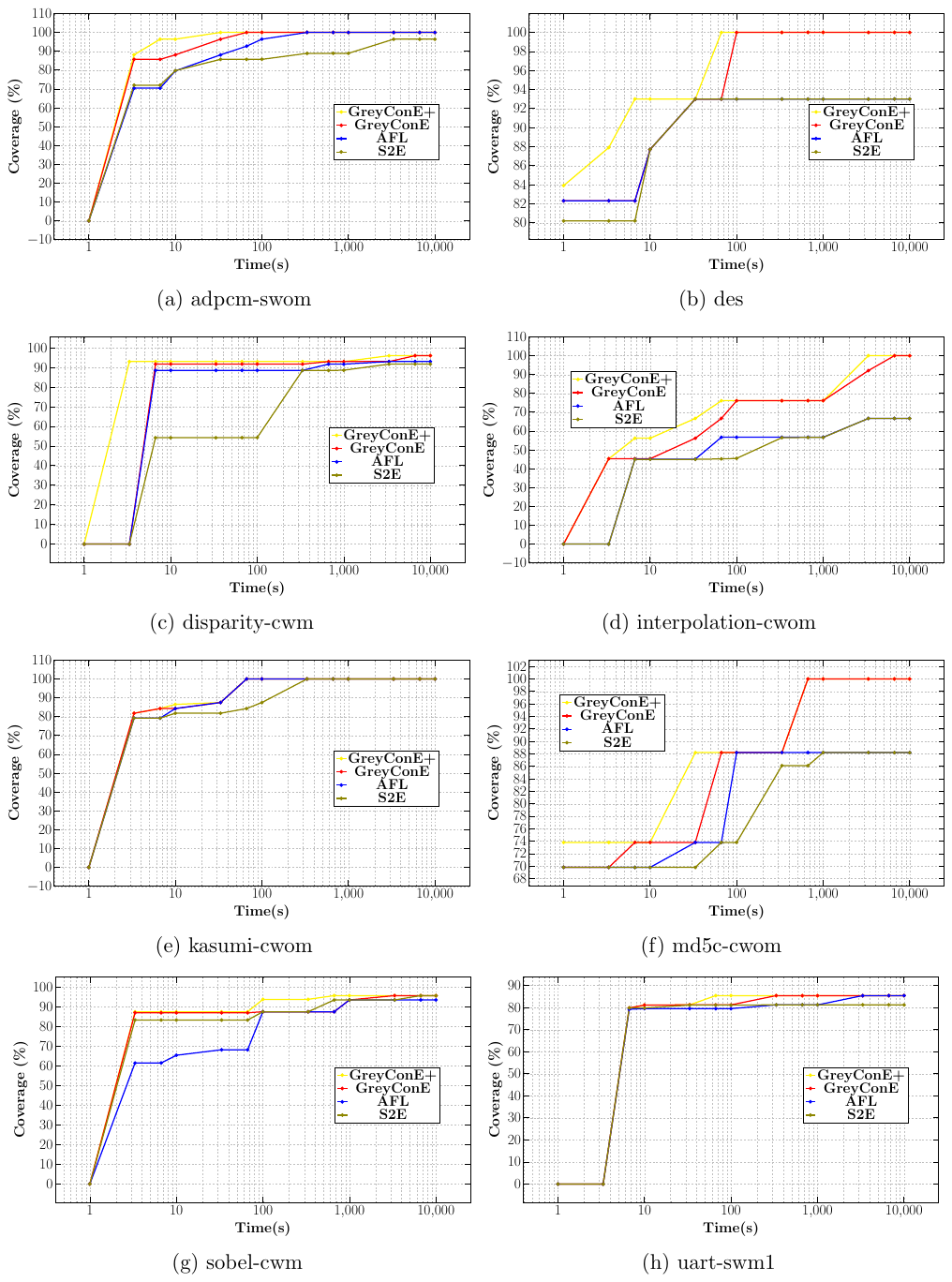}}
       \caption{Branch coverage obtained on S3C, S2C and SC benchmarks after running \emph{GreyConE+}, \emph{GreyConE}, \emph{S2E} and \emph{AFL} for two hours}
    \label{fig:coverageOfGreyConE}
\end{figure}

\begin{table*}[!tb]
  
    \centering

    \caption{Comparing Trojan detection performance of the GreyConE+ testing approach (GTD+) with baseline techniques \emph{AFL}~\cite{afl} and \emph{S2E}~\cite{s2e} and with prior works~\cite{fuzzSystemC,symbolicSystemC} and GreyConE\cite{greycone} (GTD) on on trojan infected S3CBench\cite{s3cbench_benchmark} and trojan inserted designs from S2CBench\cite{s2cbench}. Trojan types: \vl{cwm}, \ot{cwom}, \ct{swm} and \sep{swom}.}
    \setlength\tabcolsep{3pt}
    \resizebox{0.85\textwidth}{!}{
    \begin{tabular}{lccccccccccccccccc}
        \toprule
        \toprule
    \multicolumn{1}{c}{\multirow{3}{*}{\bf{Benchmarks}}} & \multicolumn{6}{c}{\bf{Test-cases generated}} & \multicolumn{6}{c}{\bf{Time taken (s)}} & \multicolumn{5}{c}{\bf{Branch Cov.(\%)}} \\ 
    \cmidrule(lr){2-7} \cmidrule(lr){8-13}  \cmidrule(lr){14-18}
   & \begin{tabular}[c]{@{}l@{}}\bf{AFL}\end{tabular} & \begin{tabular}[c]{@{}l@{}}\bf{AFL-}\\\textbf{SHT}\end{tabular} & \begin{tabular}[c]{@{}l@{}}\\\bf{S2E}\end{tabular} & \begin{tabular}[c]{@{}l@{}}\bf{SCT-}\\ \bf{HTD}\end{tabular} & \begin{tabular}[c]{@{}l@{}}\\ \bf{GTD} \end{tabular} & \begin{tabular}[c]{@{}l@{}}\\ \bf{GTD+} \end{tabular} & \bf{AFL} & \begin{tabular}[c]{@{}l@{}}\textbf{AFL-}\\ \textbf{SHT}\end{tabular} & \bf{S2E} & \begin{tabular}[c]{@{}l@{}}\bf{SCT-}\\ \bf{HTD}\end{tabular} & \begin{tabular}[c]{@{}l@{}}\\ \bf{GTD} \end{tabular} & \bf{GTD+} & \bf{AFL} & \begin{tabular}[c]{@{}l@{}}\textbf{AFL-}\\ \textbf{SHT}\end{tabular} & \bf{S2E} &  \begin{tabular}[c]{@{}l@{}}\\ \bf{GTD} \end{tabular} & \begin{tabular}[c]{@{}l@{}}\\ \bf{GTD+} \end{tabular} \\
\cline{1-18}

         \ct{adpcm-swm} & 34 & 423 & 29 & 27 & 14 & \textbf{6} & TO & 1.17 & \rt{TO} & 157 & 438 & 258 & 96.4 & 100 & 96.4 & 100 & 100 \\
         \sep{adpcm-swm} & 33 & 414 & 23 & 7 & 14 & \textbf{6} & 126.9 & 1.67 & \rt{TO} & 31 & 52 & 20 & 100 & 100 & 96.4 & 100 & 100 \\ \cmidrule(lr){1-18}
         \ot{aes-cwom} & 23 & 22 & 8 & 11 & 13 & \textbf{7} & \rt{TO} & 0.04 & \rt{TO} & 23 & 159.07 & 32 & 97.1 & 100 & 97.1 & 100 & 100  \\ \cmidrule(lr){1-18}
         \ct{decimation-swm} & 4 & NA & 5 & NA & 5 & 4 & 3196.5 & NA & 1643 & NA & 839 & \textbf{132} & 96.8 & NA & 96.8 & 96.8 & 96.8 \\
        \cmidrule(lr){1-18}
        \vl{disparity-cwm} & 20 & 36391 & 54 & NA & 66  & 65 & \rt{TO} & 63.70 & \rt{TO} & - & 3521.60 & 2686 & 93.0 & 93.2 & 91.7 & 96.0 & 96.0 \\ 
        \ot{disparity-cwom} & 38 & 36736 & 4 & NA & 62 & 57 & 3692.2 & 65.97 & \rt{TO} & NA & 3594.02 & 2319 & 96.9 & 93.8 & 91.7  & 96.9 & 96.9 \\ \cmidrule(lr){1-18}
        \ot{FFT\_fixed-cwom} & 9 & NA & 52 & NA & 16 & 12 & 1385.68 & NA & 2631 & NA & 227.47 & \textbf{161} & 96.9 & NA & 96.9 & 96.9 & 96.9 \\ \cmidrule(lr){1-18}
        \ot{filter\_FIR-cwom} & 3 & 41 & 5 & 26 & 6 & 3 & 27.8 & 0.07 & 144 & 13 & 24 & 6 & 93.8 & 93.8 & 93.8 & 93.8 & 93.8 \\ \cmidrule(lr){1-18}
         \ot{idct-cwom} & 132 & NA & 146 & NA & 110 & \textbf{97} & 36.11 & NA & 422 & NA & 138.2 & \textbf{14} & 93.8 & NA & 93.8 & 93.8 & 93.8 \\ \cmidrule(lr){1-18}
        \ot{interpolation-cwom} & 60 & 2325402 & 72 & NA & 113 & 88 & \rt{TO} & \rt{TO} & \rt{TO} & NA & 3657 & \textbf{2311} & 66.7 & 68.8 & 66.7 & 100 & 100 \\
        \ct{interpolation-swm}& 24 & 47 & 8 & NA & 3 & 3 & 10.2 & 0.16 & 295 & NA & 10.2 & 19 & 100 & 100 & 100 & 100 & 100 \\
        \sep{interpolation-swom} & 55 & 47  & 2 & NA & 3 & 3 & 10.2 & 0.16 & 216 & NA & 10.2 & 13 & 100 & 100 & 100 &  100 & 100 \\
        \cmidrule(lr){1-18}
        \ot{kasumi-cwom} & 63 & 316 & 76 & NA & 56 & 56 & 58.48 & 1.32 & 233 & NA & 58.48 & 55 & 100 & 100 & 100 & 100 & 100 \\
        \ot{kasumi-cwom} & 62 & 345 & 49 & NA & 55 & 55 & 58.55 & 1.32 & 278 & NA & 58.55 & 57 & 100 & 100 & 100 & 100 & 100 \\ \cmidrule(lr){1-18}
        \ot{md5c-cwom} & 11 & NA & 45 & NA & 10 & 10 & \rt{TO} & NA & \rt{TO} & NA & 577 & \textbf{574} & 88.2 & NA & 88.2 & 100 & 100 \\ \cmidrule(lr){1-18}
         \vl{sobel-cwm} & 98 & 10330 & 14 & NA & 57 & 46 & \rt{TO} & 4157.07 & 4892 & NA & 2537 & \textbf{682} & 93.5 & 94.2 & 95.7 & 95.7 & 95.7 \\
        \ot{sobel-cwom} & 17 & 190 & 5 & NA & 12 & 12 & 685.9 & 64.02 & 592 & NA & 612.64 & 771 & 97.8 & 98.1 & 97.8 & 97.8 & 97.8 \\
        \ct{sobel-swm} & 23 & 182 & 5 & NA & 16 & 16 & 676.8 & 58.65 & 605 & NA & 489.72 & 469 & 94.4 & 95.2 & 94.4 & 94.4 & 94.4 \\
        \cmidrule(lr){1-18}
        \ct{uart-1} & 7 & NA & 15 & 6 & 9 & \textbf{4} & 2963.6 & NA & 145 & 9 & 268 & 68 & 85.4 & NA & 81.2 & 85.4 & 85.4\\
        \ct{uart-2} & 4 & 51 & 2 & 3 & 4 & 4 & 2385.5 & 0.18 & 106 & 9 & 274 & 119 & 86.0 & 85.4 & 80.0 & 86.0 & 86.0 \\
\bottomrule
\bottomrule
\end{tabular}
\label{tab:trojanDetectionGreyConE+}
}
\end{table*}

\subsection{Trojan Detection}

In this section, we showcase the trojan detection capability of \textit{GreyConE+}. The trojan detection module confirms the identification of trojans when the rarest triggers are activated. Our experiments focus on the \textit{S3CBench}, a benchmark suite with trojans inserted in SystemC designs. To illustrate the scalability of our approach, we customize three designs from the trojan-free \textit{S2CBench} by introducing trojans in the designs. These three designs are specifically customized with \textit{cwom} trojans, incorporating characteristics from the \textit{S3CBench} suite.
To detect trojans in the 3PIPs by the CSP, we assume the availability of input-output response pairs from a functionally correct golden reference model procured from a trusted 3PVT. 
The target test cases generated by \textit{GreyConE+} are replayed on the DUT and matched with the reference output for the same set of test-cases. Trojan-infected designs will exhibit different outputs, indicating information leakage or contradictory results compared to the golden outputs. A mismatch between these two outputs implies that the design deviates from its specified functionality, revealing the presence of a trojan triggered by that specific test case.
We compare the results for trojan detection by the baseline techniques: \textit{AFL} and \textit{S2E}, state-of-the-art works: \textit{AFL-SHT}~\cite{fuzzSystemC}, \textit{SCT-HTD}~\cite{symbolicSystemC} and \textit{GreyConE}~\cite{greycone} as a trojan detection framework (\textsc{GTD}), with \textit{GreyConE+} framework for detecting trojans (\textsc{GTD+}) on the trojan infected SystemC designs, as shown in~\autoref{tab:trojanDetectionGreyConE+}. 

In this work, we compare wall-time for trojan detection by each technique. The reason being: fuzz engine takes less CPU-time for a given wall-time as it is an IO intensive process. Whereas concolic engine takes more CPU-time for a given wall-time as it forks multiple threads for test generation. For a fair comparison across a range of techniques, we choose wall-time as a metric to compare with previous works. The wall-time taken for trojan detection is presented in~\autoref{tab:trojanDetectionGreyConE+} (Column 3). TO indicates trojan is not detected within the wall-time limit of two hours. The data shows that \textit{GTD+} works better than vanilla
AFL for half of the designs and outperforms S2E for almost all the designs. GTD+ avoids expensive path exploration by concolic execution, using fuzzer generated seeds effectively reducing time for test-case generation.
\textit{GTD+} does not time-out on any benchmark however takes longer time than \textit{AFL-SHT} for most of the designs. 
\textit{GTD+ }detects the trojan in \textit{interpolation-cwom} in 2305s whereas \textit{AFL-SHT} timed out. 
\textit{AFL-SHT} employs pump mutation, format-aware trimming, and design-aware interesting number generation to optimize test-generation for trojan detection.
\textit{GTD+} works better than \textit{SCT-HTD} for benchmarks like \textit{adpcm-swom} and \textit{Filter\-FIR}. 
Also data is not available in \textit{AFL-SHT} and \textit{SCT-HTD} for all the designs of \autoref{tab:trojanDetectionGreyConE+} that we have experimented. Coverage data for \textit{SCT-HTD} are not reported in their paper. \textit{SCT-HTD} employs selective concolic execution excluding the library and pragmas and works on conditional statements and loops by forking two states for each conditions and iterations. 
The results indicates that \textit{GTD+} framework can generalize well across diverse set of benchmarks at the cost of allowing \textit{AFL} to perform aggressive mutations (rather than restricting it to a few types of mutations suitable for certain benchmarks). Also \textit{GreyConE+} avoid the path explosion problem of symbolic execution by controlled forking in symbolic loops using the fuzzer-generated test cases, thereby reaching deeper code segments without generating lots of states.
Our findings demonstrate that \textit{GTD+} efficiently detects trojans across all benchmark circuits compared to both baseline techniques and state-of-the-art methods. Particularly noteworthy are the significant reductions in required test cases and memory usage, which we discuss next.

\noindent
\textbf{Test-case Quality:} As depicted in~\autoref{tab:trojanDetectionGreyConE+}, the \textit{GTD+} framework needs fewer test-cases than both \textit{AFL} and \textit{S2E} for trojan detection and outperforms\textit{ AFL-SHT}, \textit{SCT-HTD} and GTD by detecting trojans with fewer test cases. 
This reduction in the number of test cases is due to the combined strategies of selective instrumentation and concrete-symbolic execution by \textit{GTD+}. The selective instrumentation guides and accelerates the fuzzer and the fuzzer generated input seed guides the symbolic engine to construct the execution tree along the execution path triggered by the fuzzer seed and generate test cases that only expands the tree. This leads to lower number of test-case generation to cover the trojan points faster.

It is to be noted that the number of effective test-cases reported for \textit{AFL-SHT} is as given in the paper~\cite{fuzzSystemC}, which includes the total number of test inputs tried by it till completion. The number reported for GTD+ is the final number of test cases that are fed to the Trojan Detector.  Although \textit{SCT-HTD} performs better for the design \textit{uart}, it heavily depends on the exploration strategy selected by a concolic engine to explore the search space. Upon closer examination, it becomes evident that in designs where \textit{AFL} and \textit{S2E} standalone fail to progress and experience timeouts, rendering them unable to produce test cases, \textit{GTD+} demonstrates the ability to explore additional test cases and uncover trojans. 
However, the quantity of test cases generated by \textit{GTD+} is lower compared to \textit{GTD}, primarily because \textit{GTD+} employs a faster exploration strategy achieved through selective instrumentation of interesting nodes. Thus overall \textit{GTD+} indicates good quality test-case generation.

\begin{figure}
    \centering
    \resizebox{\textwidth}{!}{
    \includegraphics{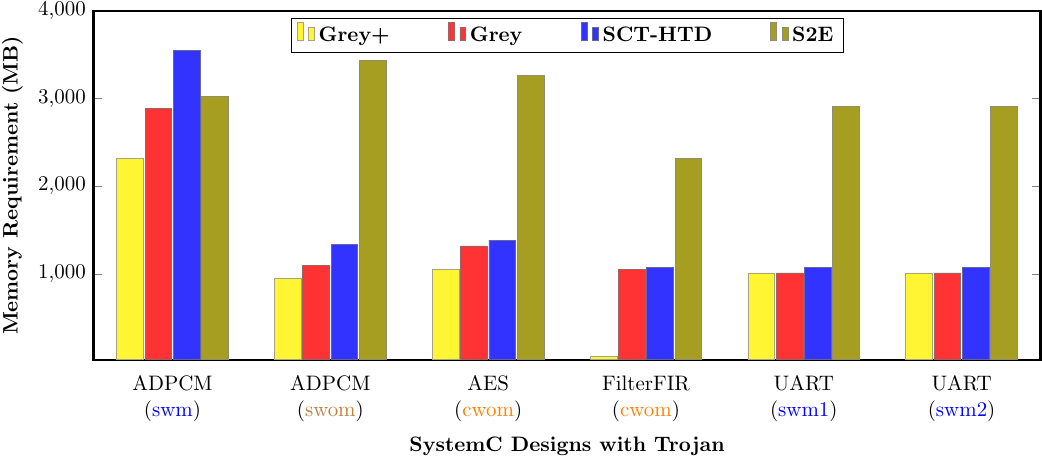}}
     \caption{Memory usage by \textit{S2E}, \textit{SCT-HTD}, \textit{GreyConE} and \textit{GreyConE+} for trojan detection}
\label{fig:greyp-memory}
\end{figure}

\noindent
\textbf{Memory Footprints:} From the~\autoref{fig:greyp-memory}, we observe that \textit{GTD+} effectively detects the trojan while maintaining reasonable memory usage.
We compare it with \textit{GTD}, \textit{SCT-HTD} as reported in~\cite{symbolicSystemC} and concolic execution engine \textit{S2E}.
\textit{GTD} performs almost similar to \textit{GTD+} when execution phases are same, whereas \textit{SCT-HTD} and \textit{S2E} exhibit higher memory consumption.
\textit{GTD+} and \textit{GTD} employ a combination of fuzz testing and concolic execution, resulting in lower memory consumption. This is attributed to the nature of fuzzing, which is I/O bound and therefore conserves memory. Conversely, SCT-HTD and S2E rely solely on symbolic execution, a memory-intensive process. Engines like \textit{S2E} allocate substantial memory for applications to run on virtual machines.
For fuzzer \textit{AFL}, the default memory configuration of 50 MB suffices for the benchmarks utilized. Notably, \textit{GTD+} completes certain designs in phase 1 ($fuzz_1$), where only \textit{AFL}'s memory requirement is necessary, as demonstrated for \textit{Filter-FIR (cwom)} in~\autoref{fig:greyp-memory}. Memory usage data for \textit{ AFL-SHT }is not reported in their paper. From the reported result one can interpret that fuzzing combined with concolic execution can lead upto $50\%$ less memory footprints compared to standalone concolic execution.

\subsection{Case Studies}
Here, we dive deep into \textit{GreyConE+}'s performance on two of the benchmark designs across two orthogonal directions: 1) Efficient coverage of rare targets and 2) Trojan detection efficiency, within a defined time  limit.\\

\noindent{\textbf{1) Case Study I: Rare Trigger Coverage Analysis}}

\begin{lstlisting}[caption={Face Detection code showing uncovered region},style=customc,label={label:face-detect}]
/* Cascade Classifier for Face Detection */
.............
tr2.x =     rectangles_array8[haar_counter]; tr2.width = rectangles_array10[haar_counter]; ?\label{line:line1}?
tr2.y =     rectangles_array9[haar_counter]; tr2.height = rectangles_array11[haar_counter]; ?\label{line:line2}?
//coord[0] - coord[8] are assigned here
if (!(tr2.x ==0&&tr2.width==0&&tr2.y==0&&tr2.height==0)&&tr2.width!=0&&tr2.height!=0) ?\label{line:branch}?//Branch 1
 {
   coord[8]=SUM1_data[pt.y+tr2.y][pt.x+tr2.x];coord[9]=SUM1_data[pt.y+tr2.y][pt.x+tr2.x+tr2.width]; 
   coord[10]=SUM1_data[pt.y+tr2.y+tr2.height][pt.x+tr2.x];coord[11]=SUM1_data[pt.y+tr2.y+tr2.height][pt.x+tr2.x+tr2.width];     
 }
 else { coord[8] = 0; coord[9] = 0; coord[10] = 0; coord[11] = 0; }
...............
      
      
      
      
\end{lstlisting}

\begin{lstlisting}[caption={Face Detection code with symbolic inputs for the uncovered condition},style=customc,label={label:face-detect-sym}]
/* Cascade Classifier for Face Detection */
.............    
//Making the input variables for tr2 symbolic for symbolic run
s2e_make_symbolic(&tr2.x, sizeof(tr2.x), "tr2.x"); ?\label{line:symb1}? //symbolic input
s2e_make_symbolic(&tr2.width, sizeof(tr2.width), "tr2.width");
s2e_make_symbolic(&tr2.y, sizeof(tr2.y), "tr2.y"); 
s2e_make_symbolic(&tr2.height, sizeof(tr2.height),"tr2.height"); ?\label{line:symb2}?
...............
      
      
      
      
\end{lstlisting}

\noindent
\textbf{{\textit{Face Detection:}} } In~\autoref{label:face-detect}, we focus on a critical segment within the cascade classifier function of Face Detection design from \emph{Rosetta} benchmark suit. This function represents a significant computational load in the face detection application. Within the listing, we draw attention to a specific branch ($Branch~1 (line~\ref{line:branch})$)  that remains elusive during random simulations. This is due to a complex conditional check, making occurrences of the associated events rare and thus reducing the likelihood of hitting the branch.
However, several other branches in the code rely on the execution of $Branch 1$. To address this, we employ \textit{GreyConE+} to enhance coverage, particularly targeting $Branch~1$ and other infrequent branches. 
We feed the list of functions and the rare branch conditions as input for selective instrumentation. Selective instrumentation is applied for $Branch~1$ of \texttt{cascadeClassifier} function and other dependent branches of \texttt{cascadeClassifier}, enabling the fuzzer to generate test cases tailored to these specific functions and conditions.
Accordingly, we modify the code as outlined in~\autoref{label:face-detect} with coverage directive functions \texttt{afl\_coverage\_interesting(u8 val, u32 id)} to prepare for the $fuzz_1$ phase. We put the parameters for \texttt{val} as our requirements for the variables of \texttt{tr2} of $Branch~1$ and the \texttt{id} parameter is overwritten by the fuzzer. 

Despite the efforts during the $fuzz_1$ stage, \textit{GreyConE+} initially fails to cover $Branch1$, although it does generate test cases that leads to execution of this branch. Within five consecutive executions when the coverage improvement ceases over $t_{theshold_f}$, \textit{GreyConE+} invokes the concolic engine S2E for further exploration. This leads to additional modifications of the code region, as depicted in~\autoref{label:face-detect-sym}, to support symbolic execution during phase2.
By employing the \texttt{make\_symbolic} function of S2E, we make the members of the structure \texttt{tr2} symbolic ($line~\ref{line:symb1}-\ref{line:symb2}$) instead of its initialization as shown in \autoref{label:face-detect} ($line~\ref{line:line1}-\ref{line:line2}$), facilitating the exploration of various execution paths corresponding to different inputs. This allows \textit{GreyConE+} to exhaustively cover all possible execution paths without surpassing the time threshold of the concolic engine.
Ultimately, although \textit{GreyConE+} invests more time in the concolic phase, it successfully achieves comprehensive coverage of rare conditions such as $Branch1$, consequently enabling the execution of other target branches. This results in the attainment of 23 out of 30 rare target coverage goals in a shorter time frame compared to \textit{GreyConE}. In contrast, AFL and S2E individually cover only 8 out of 30 branches.

\begin{figure}
    \centering
    \resizebox{\textwidth}{!}{
    \includegraphics{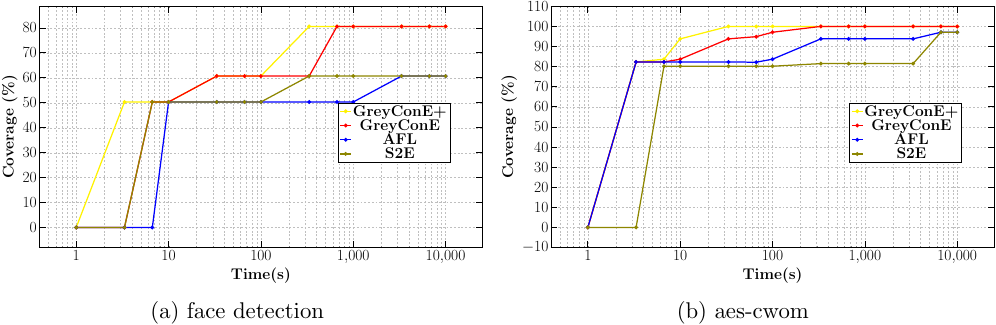}}
        \caption{Branch coverage obtained on the benchmarks from case studies after running \emph{GreyConE+}, \emph{GreyConE}, \emph{S2E} and \emph{AFL} for two hours}
    \label{fig:coverageOfGreyConE}
\end{figure}

\noindent {\textbf{2) Case Study II: Trojan Detection}}

\noindent
\textbf{{\textit{Advanced Encryptio Standard}}} (aes) is a symmetric block cipher algorithm. The plain text is 128 bits long and key can be 128, 192 or 256 bits. \emph{S3CBench} contain aes-128 bit design and the trojan type is \textit{cwom}. This trojan leaks the secret key for a specific plain-text input corrupting the encryption generating incorrect cipher-text. aes-128 performs ten rounds of repetitive operation to generate cipher-text $CT_{10}$. The trojan implementation performs an additional 'n' rounds to generate cipher-text ($CT_{10+n}$). Using \textit{GreyConE+}, we compare the generated cipher-text with the expected cipher-text to detect the presence of a Trojan. \\

 We indicate the Trojan infected \textit{aes} in~\autoref{label:aescode} ($lines~\ref{line:trigger}-\ref{line:payload}$). The Trojan trigger condition for \emph{aes} is a rare combination of input values that do not emerge in the random simulation results. The conditional path contains rare constraints and random sequences that lead to that target branch’s low probability to emerge in the random simulation results. AFL and \ac{S2E} supplied with random test-cases was unable to detect the trojan (\autoref{tab:trojanDetectionGreyConE+}) and timed-out with a branch coverage of 97.1\%. 
 The GTD+ framework passes the aes design to its fuzz engine with selective instrumentation of the trojan branch at $line~\ref{line:trigger}$ and compiled with coverage directive for the \texttt{data3[0]} value through the function \texttt{afl\_coverage\_interesting(u8 val, u32 id)}.  GTD+ at $fuzz_1$ stage produces quality seeds in less time compared to naive fuzzing. However, the fuzzer timed out at $t_{threshold_f}$ and GTD+ needed to invoke the concolic engine with the fuzzer generated seeds. 
 Since the trojan triggers depend on the input data \texttt{data3[0]} for plaintext, we define this variable as symbolic using the function \texttt{s2e\_make\_symbolic} at~\autoref{label:aescode-sym} ($line~\ref{line:symbplain}$). S2E performs concolic execution with the fuzzer generated seeds. 

\begin{lstlisting}[caption={aes - Trojan Logic}, style=customc,label={label:aescode}]
..............
for (k = 0; k < NB; k++) {
            data3[k] = data[k]; //data is the plaintext message
        }
        Computing(data, sboxs); //Encrypt data   
        /*** Output data ****/
//Trojan Triggers for a particular input present in the plaintext message 
#ifdef CWOM 
    if (data3[0] == -1460255950) ?\label{line:trigger}? //trojan trigger logic
        odata[(i * NB) + j] = (data1[i] $>>$ (j * 8));?\label{line:payload}? //trojan payload circuit
    else  odata [( i * NB ) + j ] = 0; ?\label{line:trojan-free}?//trojan free execution output 
#endif
............
\end{lstlisting}

\begin{lstlisting}[caption={aes - symbolic inputs for Trojan Logic}, style=customc,label={label:aescode-sym}]
..............
s2e_make_symbolic(&data3[0], sizeof(data3), "data3"); ?\label{line:symbplain}? //symbolic input
for (k = 1; k < NB; k++) { ?\label{line:plaintextinput}? 
            data3[k] = data[k];} //data is the plaintext message
............
\end{lstlisting}
 
 \textit{GTD+} successfully detected the Trojan in \textit{aes-cwom} in phase $fuzz_1$-$conc_1$ faster than GTD which completed at $fuzz_2$ phase. GTD+ took 26s for trojan detection which is closer to SCT-HTD which took 23, however the memory consumption SCT-HTD is more than GTD+ (\autoref{fig:greyp-memory}). AFL-SHT excels GTD+ with time for trojan detection, but needs more number of test-cases than GTD+ (\autoref{tab:trojanDetectionGreyConE+}).
\\

\label{label:exResults}

\section{Conclusion}
\label{sec:conclusion}
In this paper, we propose a rare-target test-generation framework that penetrates deeper program segments in HLS designs executed on cloud FPGAs. Our results demonstrate faster coverage of rare targets and more effective detection of trojans hidden in rare locations across a variety of benchmark designs compared to earlier methods. The complete \emph{GreyConE+} framework has the potential to significantly enhance the automated detection of security vulnerabilities in HLS designs\cite{farimah2021}.
The efficiency of GreyConE+ in terms of time, number of test cases, and memory requirements is highly desirable for use in cloud environments, where time, energy, and storage are valuable resources. However, the results of GreyConE+ depend heavily on the selection of random input stimuli and threshold parameters. In the future, we aim to address these limitations to make the framework more generalized and adaptable.
We are also investigating how trojans in RTL/gate-level netlists manifest in HLS using tools like VeriIntel2C and Verilator. Future work includes exploring input grammar-aware fuzzing and focusing more on coverage metrics such as Modified Condition/Decision Coverage (MC/DC)~\cite{codeCoverage} and path coverage.


\bibliographystyle{ACM-Reference-Format}
\bibliography{sample-base}


\end{document}